\input harvmac 
\input epsf.tex

\overfullrule=0mm

\newcount\figno
\figno=0
\def\fig#1#2#3{
\par\begingroup\parindent=0pt\leftskip=1cm\rightskip=1cm\parindent=0pt
\baselineskip=11pt
\global\advance\figno by 1
\midinsert
\epsfxsize=#3
\centerline{\epsfbox{#2}}
\vskip 12pt
{\bf Fig. \the\figno:} #1\par
\endinsert\endgroup\par
}
\def\figlabel#1{\xdef#1{\the\figno}}
\def\encadremath#1{\vbox{\hrule\hbox{\vrule\kern8pt\vbox{\kern8pt
\hbox{$\displaystyle #1$}\kern8pt}
\kern8pt\vrule}\hrule}}


\def\IR{\relax{\rm I\kern-.18em R}}
\font\cmss=cmss10 \font\cmsss=cmss10 at 7pt

\font\cmss=cmss10 \font\cmsss=cmss10 at 7pt
\def\IZ{\relax\ifmmode\mathchoice
{\hbox{\cmss Z\kern-.4em Z}}{\hbox{\cmss Z\kern-.4em Z}}
{\lower.9pt\hbox{\cmsss Z\kern-.4em Z}}
{\lower1.2pt\hbox{\cmsss Z\kern-.4em Z}}\else{\cmss Z\kern-.4em Z}\fi}
\def\IN{\relax{\rm I\kern-.18em N}}


\Title{\vbox{\hsize=3.truecm \hbox{SPhT/02-170}}}
{{\vbox {
\bigskip
\centerline{Geometrically Constrained Statistical Models}
\medskip
\centerline{on Fixed and Random Lattices:}
\medskip
\centerline{From Hard Squares to Meanders}
}}}
\bigskip
\centerline{ 
P. Di Francesco\foot{philippe@spht.saclay.cea.fr}}
\medskip
\centerline{ \it CEA-Saclay, Service de Physique Th\'eorique,}
\centerline{ \it F-91191 Gif sur Yvette Cedex, France}
\bigskip

We review various combinatorial applications of field theoretical
and matrix model approaches to equilibrium statistical
physics involving the enumeration of fixed and random lattice model
configurations. We show how the structures of the underlying lattices,
in particular their colorability properties, become relevant when we
consider hard-particles or fully-packed loop models on them.  
We show how a careful back-and-forth application of results of
two-dimensional quantum gravity and matrix models allows to predict
critical universality classes and consequently
exact asymptotics for various numbers, counting in particular hard 
object configurations on fixed or random lattices and meanders.

\Date{11/02}

\nref\HARD{J. Bouttier, P. Di Francesco and E. Guitter, {\it Critical
and tricritical hard objects on bicolourable random lattices: exact solutions}
J. Phys. A: Math. Gen. {\bf 35} (2002) 3821-3854.}
\nref\ASY{P. Di Francesco, O. Golinelli and E. Guitter,
{\it Meanders: exact asymptotics}, Nucl.Phys. B570 (2000) 699-712.}
\nref\FOLCO{P. Di Francesco, {\it Folding and Coloring Problems in Mathematics and Physics},
Bulletin of the AMS, Vol. {\bf 37}, No. {\bf 3} (2000) 251-307.}
\nref\GF{D. Gaunt and M. Fisher, 
{\it Hard-Sphere Lattice Gases.I.Plane-Square Lattice}, J. Chem. Phys. 
{\bf 43} (1965) 2840-2863.}
\nref\RUN{L. Runnels, L. Combs and J. Salvant, {\it Exact Finite 
Methods of Lattice Statistics. II. Honeycomb-Lattice Gas of Hard Molecules}, 
J. Chem. Phys. {\bf 47} (1967) 4015-4020.}
\nref\BAXHS{R. J. Baxter, I. G. Enting and S.K. Tsang, {\it Hard Square Lattice Gas},
J. Stat. Phys. {\bf 22} (1980) 465-489.}
\nref\CFT{P. Di Francesco, P. Mathieu and D. S\'en\'echal, {\it Conformal Field Theory},
Graduate Texts in Contemporary Physics,
Springer (1996) 1-890 (1st ed.) and Springer (1999) 1-890 (2nd ed.).}
\nref\CIZ{A. Cappelli, C. Itzykson and J.-B. Zuber, {\it The A-D-E classification of minimal and
$A_1^{(1)}$ conformal invariant theories}, Comm. Math. Phys. {\bf 113} (1997), 1-26.}
\nref\DGZ{P. Di Francesco, P. Ginsparg
and J. Zinn--Justin, {\it 2D Gravity and Random Matrices},
Physics Reports {\bf 254} (1995) 1-131.}
\nref\BAXHH{R. J. Baxter, {\it Hard Hexagons: Exact Solution}, J. Phys. {\bf A 13}
(1980) L61-L70; R. J. Baxter and S.K. Tsang, {\it Entropy of Hard Hexagons},
J. Phys. {\bf A 13} (1980) 1023-1030; see also
R. J. Baxter, {\it Exactly Solved Models in Statistical Mechanics},
Academic Press, London (1984).}
\nref\BAX{R. J. Baxter, {\it Planar Lattice Gases with Nearest-neighbour
Exclusion}, Annals of Combin. No. {\bf 3} (1999) 191-203 preprint cond-mat/9811264.}
\nref\KF{D. Kurze and M. Fisher, {\it Yang-Lee Edge Singularities
at High Temperatures}, Phys. Rev. {\bf B20} (1979) 2785-2796.}
\nref\CARTWO{J. Cardy, {\it Conformal Invariance and the Yang-Lee Edge
Singularity in Two Dimensions}, 
Phys. Rev. Lett. {\bf 54}, No. 13 (1985) 1354-1356.}
\nref\MMC{P. Di Francesco, {\it Matrix Model Combinatorics: Applications to Folding and
Coloring}, Bleher and Its Eds., MSRI Publications Vol. {\bf 40} 111-170,
Cambridge University Press (2001), math-ph/9911002.}
\nref\EY{B. Eynard, {\it Random Matrices}, Saclay Lecture Notes (2000),
\hfill\break 
http://www-spht.cea.fr/lectures\_notes.shtml }
\nref\ITZ{C. Itzykson and J.-B. Zuber, {\it The planar
approximation II}, J. Math. Phys. {\bf 21} (1980) 411.}
\nref\HARDOB{J. Bouttier, P. Di Francesco and E. Guitter,
{\it Combinatorics of Hard Particles on Planar Graphs},
Saclay preprint t02/160 and cond-mat/0211168 (2002).}
\nref\CONCUR{M. Bousquet-M\'elou and G. Schaeffer, {\it The degree
distribution in bipartite planar maps: applications to the Ising model},
preprint math.CO/0211070 (2002).}
\nref\TUT{W. Tutte, {\it A Census of Planar Maps}, Canad. Jour. of Math. 
{\bf 15} (1963) 249-271.}
\nref\SCH{G. Schaeffer, {\it Bijective census and random
generation of Eulerian planar maps}, Electronic
Journal of Combinatorics, vol. {\bf 4} (1997) R20; {\it Conjugaison d'arbres
et cartes combinatoires al\'eatoires} PhD Thesis, Universit\'e
Bordeaux I (1998).}
\nref\KPZ{V.G. Knizhnik, A.M. Polyakov and A.B. Zamolodchikov, {\it Fractal Structure of
2D Quantum Gravity}, Mod. Phys. Lett.
{\bf A3} (1988) 819-826; F. David, {\it Conformal Field Theories Coupled to 2D Gravity in the
Conformal Gauge}, Mod. Phys. Lett. {\bf A3} (1988) 1651-1656; J.
Distler and H. Kawai, {\it Conformal Field Theory and 2D Quantum Gravity}, 
Nucl. Phys. {\bf B321} (1989) 509-527.}
\nref\RECT{P. Di Francesco, {\it Rectangular Matrix Models and Combinatorics of Colored
Graphs}, preprints SPhT/02-096 and cond-mat/0208037 (2002), to appear in Nucl. Phys. {\bf B}.}
\nref\SL{A. Sainte-Lagu\"e,
{\it Avec des nombres et des lignes (R\'ecr\'eations Math\'ematiques)},
Vuibert, Paris (1937).}
\nref\TOU{J. Touchard, {\it Contributions \`a l'\'etude du probl\`eme
des timbres poste}, Canad. J. Math. {\bf 2} (1950) 385-398;
W. Lunnon, {\it A map--folding problem},
Math. of Computation {\bf 22} (1968) 193-199.}
\nref\DGG{P. Di Francesco, O. Golinelli and E. Guitter, {\it Meander,
folding and arch statistics}, Mathl. Comput. Modelling {\bf 26} (1997) 97-147.}
\nref\ARNO{V. Arnold, {\it The branched covering of $CP_2 \to S_4$,
hyperbolicity and projective topology},
Siberian Math. Jour. {\bf 29} (1988) 717-726.}
\nref\KOSMO{K.H. Ko, L. Smolinsky, {\it A combinatorial matrix in
$3$-manifold theory}, Pacific. J. Math {\bf 149} (1991) 319-336.}
\nref\HMRT{K. Hoffman, K. Mehlhorn, P. Rosenstiehl and R. Tarjan, {\it
Sorting Jordan sequences in linear time using level-linked search
trees}, Information and Control {\bf 68} (1986) 170-184.}
\nref\TLM{P. Di Francesco, O. Golinelli and E. Guitter,
{\it Meanders and the Temperley-Lieb Algebra},
Commun. Math. Phys. {\bf 186} (1997), 1-59.}
\nref\BACH{R. Bacher, {\it Meander Algebras},
pr\'epublication de l'Institut Fourier n$^o$ $478$ (1999).}
\nref\LZ{S. Lando and A. Zvonkin, {\it Plane and Projective Meanders},
Theor. Comp.  Science {\bf 117} (1993) 227-241, and {\it Meanders},
Selecta Math. Sov. {\bf 11} (1992) 117-144.}
\nref\NOUS{P. Di Francesco, O. .Golinelli and E. Guitter, {\it Meanders:
a direct enumeration approach}, Nucl. Phys. {\bf B 482} [FS] (1996) 497-535.}
\nref\GOL{O. Golinelli, {\it A Monte-Carlo study of meanders},
Eur. Phys. J. {\bf B14} (2000) 145-155.}
\nref\MAK{Y. Makeenko {\it Strings, Matrix Models, and Meanders},
Nucl.Phys.Proc.Suppl. {\bf 49} (1996) 226-237;
G. Semenoff and R. Szabo {\it Fermionic Matrix Models} Int.J.Mod.Phys.
{\bf A12} (1997) 2135-2292.}
\nref\NIEN{B.\ Nienhuis in {\it Phase Transitions and Critical Phenomena},
Vol.\ 11, eds.\ C.\ Domb and J.L.\ Lebowitz, Academic Press 1987.}
\nref\BN{ H.W.J.\ Bl\"{o}te and B.\ Nienhuis, Phys.\ Rev.
Lett.\ 72 (1994) 1372.}
\nref\BSY{M.T.\ Batchelor, J.\ Suzuki and C.M.\ Yung,
Phys.\ Rev.\ Lett.\ 73 (1994) 2646, cond-mat/9408083. }
\nref\GDF{P. Di Francesco and E. Guitter, {\it
Entropy of Folding of the Triangular Lattice}, Europhys. Lett. {\bf 26} (1994) 455.}
\nref\BAXTRI{R. Baxter, {\it Colorings of a hexagonal lattice}, J. Math. Phys. {\bf 11}
(1970) 784-789.}
\nref\JACO{J. Jacobsen and J. Kondev, {\it Field theory of compact polymers
on the square lattice}, Nucl. Phys. {\bf B 532} [FS], (1998) 635-688,
{\it Transition from the compact to the dense phase of two-dimensional polymers},
J. Stat. Phys. {\bf 96}, (1999) 21-48.}
\nref\DGK{P. Di Francesco, E. Guitter and C. Kristjansen, {\it Fully Packed 
O(n=1) Model on Random Eulerian Triangulations}, Nucl.Phys. B549 (1999) 657-667.}
\nref\KZJ{V. Kazakov and P. Zinn-Justin, {\it Two-Matrix model with ABAB interaction},
Nucl.Phys. {\bf B546} (1999) 647-668.}
\nref\JEN{I. Jensen, {\it Enumerations of Plane Meanders}, preprint
cond-mat/9910313 (1999).}
\nref\DGJ{P. Di Francesco, E. Guitter and J. Jacobsen, {\it Exact Meander Asymptotics: a
Numerical Check}, Nucl.Phys. B580 (2000) 757-795.}
\nref\POLEM{I. Jensen and A. Guttmann, {\it Critical exponents of plane meanders},
J. Phys. {\bf A33} (2000) L187; I. Jensen, {\it A transfer matrix approach to the enumeration of
plane meanders}, preprint cond-mat/0008178 (2000).}
\nref\PLA{J. Bouttier, P. Di Francesco and E. Guitter, {\it Counting colored random
triangulations}, Nucl. Phys. {\bf B641[FS]} (2002) 519-532;
{\it Census of Planar Maps: From the One-Matrix Model
Solution  to a Combinatorial Proof}, Nucl.Phys. {\bf B645} (2002) 477-499.}
\nref\GKN{E. Guitter, C. Kristjansen, and J. Nielsen, {\it Hamiltonian Cycles 
on Random Eulerian Triangulations}, Nucl.Phys. B546 (1999) 731.}

\newsec{Introduction}

Physics provides a source of alternative approaches to well-posed mathematical
problems, allowing sometimes for less rigorous but more predictive results,
such as those discussed in these notes. The reason is that the tools employed
are often somewhat ill-defined from a purely mathematical point of view, although
they are often validated by experiment. In every physicist's toolbox, field theory
is certainly the most powerful one, as it basically allows for an effective
description of the essential behavior and properties of the situations it 
models. 

Our main concern here will be with solving well-posed combinatorial problems
by means of physical approaches and reasoning. This type of approach must have
a sort of exotic flavor when observed from a purely mathematical point of view.
The very principles it relies on borrow from the developments of field theory,
such as renormalization group techniques applied to critical phenomena.
Instead of trying to be completely exhaustive, which would go far beyond the
scope of the present notes, we will rather try to extract the essentials and main
lessons of the field-theoretical approaches. We will also present some alternative
techniques such as matrix integrals allowing to address some of 
these combinatorial problems.

The questions we will be considering here deal with fixed or random statistical
two-dimensional lattice models, used to describe two-dimensional interfaces as well as 
discrete two-dimensional quantum gravity (2DQG). 
Combinatorially, these are simply problems of enumeration of configurations
involving either decorated lattices or decorated graphs. 
More precisely, we will discuss here two classes of models
for which the details of the underlying lattice are important: these models
``feel" strongly their underlying discretized space. 
They are the hard-particle and fully-packed loop
models. We will be mainly interested in understanding the critical behavior
of these systems, namely in locating phase transitions in the space of parameters
where thermodynamic quantities, characteristic of the collective behavior
of the system at large size, become singular. As we will explain, these singularities
fall into ``universality classes" described by various field theories, and
characterized by critical exponents governing their
algebraic singularities. 

The first theory describes objects (particles) occupying the vertices of the
lattice, with a nearest neighbor exclusion constraint that no two vertices
connected by an edge may be simultaneously occupied. Despite the local
character of this constraint, it allows for the particles to ``feel" the 
details of the lattice. When for instance we increase the density of occupation
up to its maximum, particles will try to occupy every other vertex in the
lattice, thus will feel for instance whether it is bipartite or not. 

The second theory describes loops occupying edges of the lattice, with the additional
constraint that every vertex is visited by a loop. Again, 
the bipartite (or not) character of the lattice will prove crucial in the discussion of
the critical behavior of the system.

While solving the first theory will eventually lead to the exact enumeration of random
planar graphs with hard-particles on them, solving the second one will lead to an
asymptotic solution
of the so-called ``meander" problem of enumeration of the topologically
inequivalent configurations of a closed road crossing a river through a given 
number of bridges. The materials collected in this review article include in
particular Refs. \HARD\ and \ASY, where random lattice hard-particle and meander 
problems were first solved (see also the review article \FOLCO).

The paper is organized as follows. In Sect. 2, we 
collect a few definitions and results regarding the thermodynamics of 
two-dimensional statistical lattice models on fixed and random lattices,
including in particular field-theoretical ones.
Sect. 3 is devoted to the study of hard particles on various random lattices,
in the form of planar graphs with fixed valence. We show in particular how another
favorite physicist's tool, the matrix integral, may be used to completely solve
various enumeration problems involving decorated graphs. When applied to hard particles,
these allow for a complete understanding of the crystallization transition, namely
a critical value of the density of occupation below which the system is fluid
(particles occupy vertices quasi-randomly) and beyond which a crystal of particles
forms. We show however that such a transition only exists on (fixed or random)
lattices with a colorability property allowing for the existence of maximally
occupied sublattices. The critical behavior is then traced back directly to 
the existence and structure of these sublattices. We show in particular that for
bipartite lattices, the critical transition lies in the universality class of the 
critical 2D Ising model, thus proving on a random lattice some old fixed lattice
conjectures concerning hard squares or hard triangles \GF\ \RUN\ \BAXHS. 
In Sect. 4, we address fully-packed loop models on fixed and random lattices. After recalling
some known results in the fixed lattice case, we move to random lattices.
Remarkably, just like in the hard-particle case, we find that according to
whether the random lattices are bipartite or not, the critical universality
class predicted is different. By suitably following a two-flavor fully-packed loop model
from its fixed lattice version to its random lattice one, we arrive at the identification
of the universality class for meanders, which allows in particular to predict
the meander configuration exponent $\alpha={29+\sqrt{145}\over 12}$, governing
the large $n$ asymptotics of the meander number with $2n$ bridges $M_{2n}\sim g_c^{-2n}/n^\alpha$.
We gather a few concluding remarks in Sect. 5, as well as another simple prediction,
as a challenge for mathematicians to prove.

\newsec{2D Quantum Gravity and Asymptotic Graph Combinatorics}

\subsec{2D Lattice Models}

The archetypical systems studied in 2D equilibrium statistical mechanics are
lattice models, whose configurations are defined as maps $\sigma$ from,
say the vertices $i$ of a 2D lattice $\cal L$ (or more precisely a finite
connected subset thereof say of rectangular shape of size $L\times T$, denoted
by ${\cal L}_{L,T}$, and with definite, say periodic boundary conditions)
to a target space $\cal T$ usually real. The model is further defined
through an energy functional $E(\sigma)$, and each configuration is
attached a statistical weight proportional to $e^{-\beta E(\sigma)}$,
where $\beta=1/(k_B T)$, $k_B$ the Boltzmann constant and $T$ the temperature. 
One is usually interested in the properties of global functions such as
the partition function
\eqn\pf{ Z_{L,T}= \sum_\sigma e^{-\beta E(\sigma)} }
where the sum extends over the set of maps from ${\cal L}_{L,T}$ to $\cal T$. 
This is precisely the normalization of the probability $p(\sigma)=e^{-\beta
E(\sigma)}/Z_{L,T}$ of the configuration $\sigma$.
This probability weight allows to answer various questions regarding
correlations in the system, through the corresponding expectation value
denoted generically by $\langle ...\rangle=\sum_\sigma ... p(\sigma)$.

As a simple illustration, the Ising model has ${\cal T}=\{-1,+1\}$, and the images
$\sigma(i)$ are used for instance to describe the spins (magnetic moments) in metals.
The functional energy usually incorporates some information on interactions
within the system or with some external fields or forces. For the Ising
model, it reads $E(\sigma)=-J \sum_{(i,j)} \sigma(i) \sigma(j) -H\sum_i \sigma(i)$,
where the first sum extends over nearest neighboring pairs of vertices
$(i,j)$ and expresses
ferro-- ($J>0$) or antiferro-- ($J<0$) magnetic interactions, while the second sum
expresses the coupling of the system to an external magnetic field $H$. 
In general, the energy functional depends on external parameters (such as
$\beta J$ and $\beta H$ in the Ising case).

The archetypical questions one tries to answer involve the thermodynamic limit
in which the system becomes large, say $L,T\to \infty$ with $L/T$ finite, 
and on its dependence on the external parameters. In this limit, thermodynamic
functions such as $Z=\lim (Z_{L,T})^{1\over LT}$ may develop singularities 
when the parameters approach
some critical values, corresponding to phase transitions and critical phenomena.
Of particular interest are those in which a divergence occurs for the length
characteristic of the effective range of interactions, the {\it correlation length}.
The corresponding critical singularities fall into so-called universality classes, 
characterized by scaling exponents which govern the leading algebraic behavior
of the singular part of the thermodynamic functions of the system. For instance,
in the ferromagnetic
Ising model with vanishing magnetic field, one defines a ``thermal" exponent
$\alpha$, characteristic of the leading singularity of the free energy 
$F={\rm Log}\, Z$ when the temperature $T$ approaches a critical value $T_c$:
\eqn\singising{ F\vert_{\rm sing} \propto |T-T_c|^{2-\alpha} }
Similarly, at $T=T_c$, we also get a singular behavior for $F$ when we switch on
a small magnetic field $H$, with
\eqn\singH{ F\vert_{\rm sing} \propto |H|^{2-\beta} }
which defines the ``magnetic susceptibility" exponent $\beta$.
The exponents $\alpha,\beta$ are expected to be more universal than the actual value of the critical
temperature $T_c$: indeed, one finds for instance that $T_c$ depends on
the precise type of lattice on which the model is defined, while $\alpha,\beta$ only depend on
its dimensionality. 

In two dimensions, it has been shown that under some assumption of locality
of the interactions, critical singularities corresponding to the divergence
of the correlation length 
fall into universality classes described by two-dimensional conformal field
theories (CFT), upon taking the continuum critical limit of the models, in which we let
at the same time the size of the system tend to infinity and the lattice spacing tend to
zero while parameters tend to critical values, 
so that the lattice becomes the two-dimensional plane (or a higher genus Riemann
surface according to boundary conditions), and configuration maps become fields.
Conformal invariance of the resulting field theory is directly linked to the divergence of
the correlation length which implies in particular local scale invariance of the theory.
Two-dimensional CFT's have been extensively studied and partly classified, providing
us with a sort of Mendeleiev table of universality classes
of critical phenomena in two dimensions, in the form of a list of critical exponents
(see the text \CFT\ and references therein).
More precisely, CFT's are characterized by their central charge $c_m=1-6/(m(m+1))$
and the conformal dimensions of their operators $h_{r,s}=(((m+1)r-ms)^2-1)/(4m(m+1))$
respectively the central extension and highest weights of the corresponding representations
of the Virasoro algebra, $m$ some complex parameter, and $r,s$ some
positive integers, further restricted to a finite set in the case of minimal models,
where $m$ is rational. 
The conformal dimensions govern the fall-off with distance of correlation
functions. For instance we have for a conformal operator $\Phi_h$ of dimension $h$:
$\langle \Phi_h(z,{\bar z}) \Phi_h(0,0) \rangle = 1/|z|^{4h}$.
Upon identifying the operators of the theory as generating
perturbations of the various parameters away from their critical values, one may
relate the corresponding critical exponents to their conformal dimension
through\foot{This is done by dimensional analysis.
Writing the corresponding perturbation of the conformal
action functional as $(t-t_c)\int d^2z \Phi_h(z,{\bar z})$, and performing a local
rescaling $z\to \lambda z$, we find that $(t-t_c)\sim \lambda^{2h-2}$ as the action is
dimensionless. The thermodynamic free energy $F=\lim {1\over LT}{\rm Log}\, Z_{L,T}$
however is per unit of area, hence 
$(t-t_c)^{2-\alpha}\sim \lambda^{-2}$, hence $\alpha=1/(1-h)$.}  
\eqn\confexp{\alpha={1-2h\over 1-h}}
For instance, the critical 2D Ising model corresponds to a CFT with $m=3$, i.e. $c=1/2$ 
and the thermal operator associated with perturbations in the temperature has
$h_{1,3}=1/2$, so that the thermal exponent is $\alpha=0$ (in this case, the singularity
is weaker and involves the logarithm of $|T-T_c|$).
Besides, the magnetic ``spin" operator, the continuous version of the map $\sigma$,
has conformal dimension $h_{2,2}=1/16$, which leads to the magnetic
susceptibility exponent $\beta=14/15$.

A number of classifications of CFT's are now available, the most remarkable of which
is probably that of ``minimal models" with central charges $c<1$, namely CFT's with a finite
number of conformal primary fields \CIZ. They turn out
to have central charges $c=c(p,q)\equiv 1-6(p-q)^2/(p q)$, indexed by two 
coprime positive integers $p$ and $q$.
The constraint of unitarity (positivity of correlators, and in particular positivity
of the probabilities $p(\sigma)$) further restricts $q$ and $p$
to obey $|p-q|=1$. In this language, the Ising model corresponds to $p=3$, $q=4$.

\subsec{2D Random Lattice Models}

\fig{Schematic representation of how we go from a 2D statistical lattice model
(defined here on a finite portion of the triangular lattice) to a random lattice
model, in which the lattice is replaced by arbitrary tessellations of space
(here a random triangulation tessellating a genus one surface).}{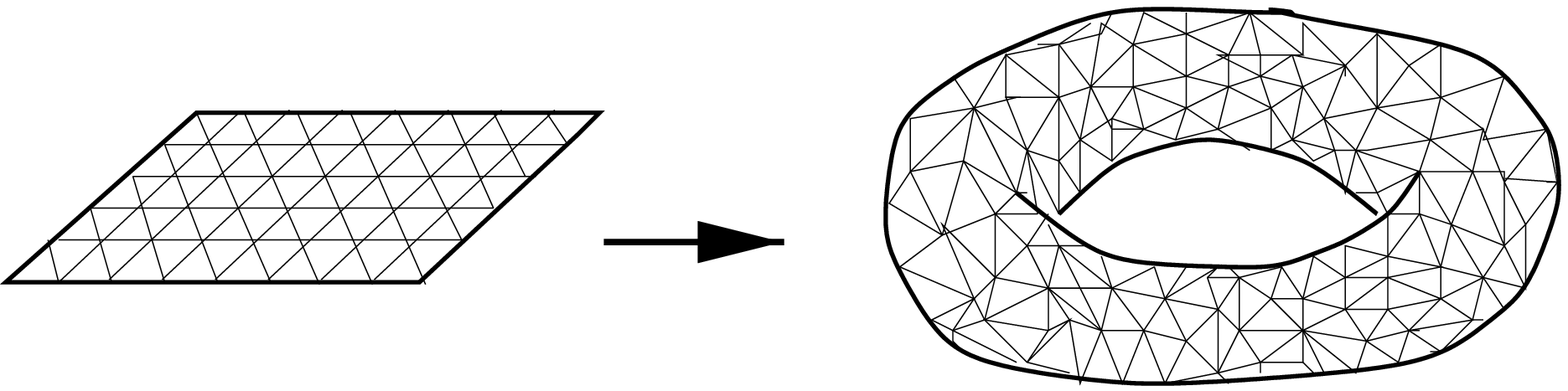}{10.cm}
\figlabel\gravite

In this work, we consider random lattice models, defined in precisely the same way as
ordinary lattice models, but by replacing the underlying lattice by a somewhat arbitrary
possibly disconnected tessellation of the plane (or higher genus surface).
The tessellation becomes therefore part of the configuration to be summed over
(see Fig. \gravite\ for an illustration).
This was introduced as a discrete model of 2D quantum gravity (2DQG), 
namely to describe the coupling of some
statistical ``matter" models to the quantum fluctuations of the underlying ``space",
represented by random lattices or tessellations (see e.g. Ref. \DGZ\ for a review
and more references).
For each tessellation $\Theta$, we associate a statistical weight directly borrowed
from the Einstein action in 2D, namely a weight $N^{\chi(\Theta)} g^{A(\Theta)}$
where $\chi$ and $A$ are respectively the Euler characteristic and area of the
tessellation (respectively measured via $\chi=\#$faces$-\#$edges$+\#$vertices and
$A=\#$tiles), and where $N$ and $g$ are the discrete counterparts of the Newton constant
and cosmological constant. 
For each tessellation $\Theta$ together with a spin configuration
$\sigma:\Theta\to {\cal T}$ on it,
we also have a weight $e^{-\beta E(\sigma)}$ and we must divide by the 
order of the automorphism group $|Aut(\Theta,\sigma)|$. 
This leads for instance to the discretized
partition function of any 2D statistical lattice model coupled to 2D quantum gravity
\eqn\twodqgpf{ Z=\sum_{{\rm tessel.}\ \Theta} 
N^{\chi(\Theta)} g^{A(\Theta)} \sum_{{\rm config.}\atop \sigma:\Theta\to {\cal T}}
e^{-\beta E(\sigma)} /|Aut(\Theta,\sigma)| }
Note that the parameter $N$ allows to isolate the contributions of tessellations
of fixed genus. In the following we will be mainly interested in the genus zero
contributions, obtained by letting $N\to \infty$.
Note also that the free energy $F={\rm Log}\, Z$ selects only the connected tessellations
in the sum \twodqgpf.

Like in the fixed lattice case, we are interested in the thermodynamic limit of the system,
in which say the average area or some related quantity diverges, ensuring that the
dominant contributions to $Z$ e.g. come from large tessellations. This is guaranteed
in general by the existence of a critical value $g_c$ of the cosmological constant $g$
at which such singularities take place. This value is a priori a function of the type
of random lattices we sum over as well as of the various matter parameters. We may now
attain interesting critical points by also letting the matter parameters approach critical
values, a priori distinct from those on fixed lattices. The result is well described
by the coupling of the corresponding CFT's with quantum fluctuations of space, namely
by letting the metric of the underlying 2D space fluctuate. Such fluctuations may be
represented in the conformal gauge by yet another field theory, the Liouville field theory,
which is coupled to the matter CFT. This field theoretical setting allows for a complete
understanding of the various critical exponents occurring at these critical points
\KPZ.
For instance, one defines the (genus zero) string susceptibility exponent $\gamma_{str}$    
as the exponent $\alpha$ associated to the cosmological constant singularity, namely
by writing the singularity of the free energy as
\eqn\frensing{ F\vert_{\rm sing} \propto (g_c-g)^{2-\gamma_{str}} }
In the case of coupling of a matter theory with central charge $c$ to 2DQG, one has the
exact relation \KPZ\
\eqn\kpz{ \gamma_{str}={c-1-\sqrt{(1-c)(25-c)} \over 12} }
In the case of ``pure gravity", namely when the matter is trivial and has $c=0$,
we get $\gamma_{str}=-1/2$, while for the critical Ising model with $c=1/2$ we 
have $\gamma_{str}=-1/3$. 
 
Upon coupling to gravity the operators of the CFT ($\Phi_h(z,{\bar z})$) 
get ``dressed" by gravity ($\Phi_h \to {\tilde \Phi_h}\equiv \Psi_\Delta$) and acquire
dressed dimensions $\Delta$, given similarly by \KPZ\
\eqn\kpzdim{\Delta={\sqrt{1-c+24 h}-\sqrt{1-c} \over \sqrt{25-c}-\sqrt{1-c}} } 
As opposed to the fixed lattice case, where conformal dimensions govern
the fall-off of correlation functions of operators with distance, the dressed operators
of quantum gravity do not feel distances, as their position is integrated over
the surfaces, but rather only feel changes of area at fixed genus. More precisely
the general genus zero correlators behave in the vicinity of $g_c$ \KPZ\ as
\eqn\corrgra{ \langle \Psi_{\Delta_1} \Psi_{\Delta_2} ... \Psi_{\Delta_n} \rangle
\sim (g_c-g)^{2-\gamma_{str}+\sum_{1\leq i\leq n} (\Delta_i-1)} }

These results may be easily translated into the large (but fixed) area $A$ behavior
of the various thermodynamic quantities, upon performing a Laplace transform,
which selects the coefficient of $g^A$ in the various expansions. Let $F_A$
denote the partition function for connected tessellations of genus zero and area $A$, 
we have
\eqn\largA{ F_A \sim {g_c^{-A} \over A^{3 -\gamma_{str}}} }
while if $\langle ...\rangle_A$ denotes  any genus zero correlator at fixed area, we have
\eqn\corrA{  \langle \Psi_{\Delta_1} \Psi_{\Delta_2} ... \Psi_{\Delta_n} \rangle_A
\sim {g_c^{-A} \over A^{3-\gamma_{str}+\sum_{1\leq i\leq n} (\Delta_i-1)} } }
In \KPZ, all these formulas were also generalized to higher genus as well.

As an illustration of eqn. \largA, recall that the number of quadrangulations of the
sphere with $A$ square tiles and with a marked edge 
(dually equal to the number of rooted tetravalent 
planar maps with $A$ vertices) reads \TUT\ 
\eqn\quadA{ N_A = {3^A\over 2(A+1)(A+2)} {2A \choose A} \sim {12^A \over A^{5/2}} }
Noting that the rooting simply amounts to $N_A \propto A F_A$,
the asymptotics \quadA\ correspond to $g_c=1/12$ and $\gamma_{str}=-1/2$, hence $c=0$. 
This is one of the various ways to attain the universality class of pure gravity, namely 
by summing over bare tessellations without matter on them.

So far so good, basically all asymptotic combinatorics
problems involving planar (or more generally fixed genus) graphs seem to be solved 
via eqns. \largA\ and \corrA\ or their higher genus generalizations, provided
one is able to identify the central charge $c$ of the underlying CFT. This latter
step however may prove to be quite involved.
In fact, the aim of this note is to clarify a large class of cases for which the naive
application of these formulas would lead to the wrong result. These are statistical models
whose definition strongly relies on the structure of the underlying (fixed or
random) lattice. This is the case for hard particles, as well as for fully-packed
loops, the two main subjects of this note.

\newsec{Hard Particles}

The first class of problems we will be discussing is that of hard particles 
occupying or not the
vertices of the fixed or random lattices, and subject to the ``hardness" or
particle-exclusion constraint that no two adjacent vertices may be simultaneously
occupied. One usually attaches an activity $z$ per occupied vertex.
In this section, we first recall the fixed lattice results, and display the
exact or predicted critical behavior of the model. Generically it undergoes
a crystallization transition between a disordered phase of low-density
of occupation and a crystalline phase of maximal occupancy. We next turn
to the exact solution of the same model on some random lattice, only to
discover that the crystalline transition is wiped out by the sum over
lattices. To solve this puzzle, we simply notice that the transition
can take place only if the
random lattices allow for the existence of some generic crystalline configurations. We
finally show that the model, when defined on random {\it bipartite}
lattices, has a crystalline transition in the universality class of the 
two-dimensional critical Ising model coupled to 2DQG.  

\subsec{Fixed lattice results and conjectures}

\fig{Typical configurations of hard particles on the (a) hexagonal, (b) square and (c) 
triangular lattices. We have also represented in grey the excluded area translating
the nearest neighbor exclusion into an non-overlapping constraint of tiles. Their
respective shapes are triangles, squares and hexagons, hence the names 
hard triangles, squares or hexagons.}{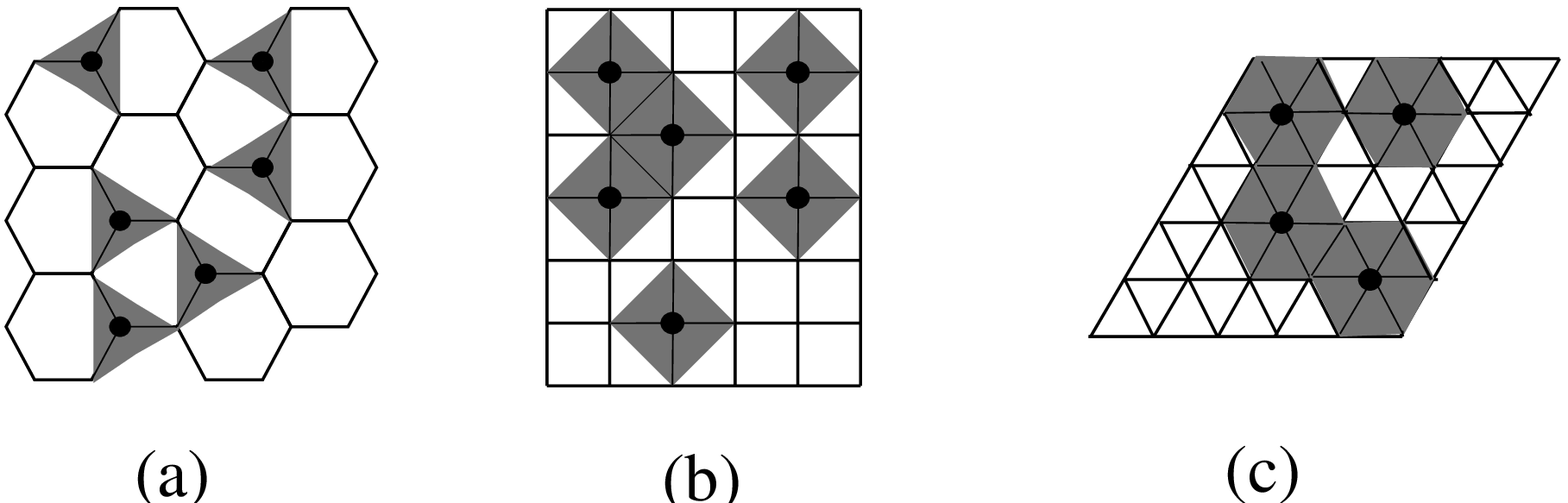}{12.cm}
\figlabel\trisquahex

Hard particle models have been extensively studied in the physics literature,
under various names: hard-core lattice gas, nearest-neighbor exclusion models,
etc ... \GF\ \RUN\ \BAXHS\ \BAXHH\ \BAX.
Among the host of results and conjectures, let us mention the three cases
of the hexagonal, square and triangular lattices.
The corresponding models are also often referred to as hard triangle, square or hexagon
models, as in the dual picture an occupied vertex may be replaced by some tile whose
shape transforms the hard-particle constraint into a no-overlap condition
between tiles (see Fig. \trisquahex\ for an illustration in the hexagonal (a),
square (b) and triangular (c) cases, respectively corresponding to hard triangles, squares and
hexagons). 
In the three cases, one expects in the thermodynamic limit
a crystallization transition to take place 
at some critical value $z=z_+>0$ between a low-$z$ disordered phase and a high-$z$
crystalline phase in which, when $z\to \infty$, only one of a few available
``groundstates" are realized, namely some maximally occupied configurations,
in which the particles occupy a sublattice of the original one.
For the hexagonal and square cases, where the lattice is bipartite, there are
exactly {\it two} such crystalline groundstates, as opposed to the triangular
case, where there are {\it three}. 
In addition, one expects as well some ``non-physical" critical point at some
other value $z=z_-<0$, independent of the precise details of the lattice, and generically
described by the Yang-Lee edge singularity \KF\ 
(of the Ising model in the presence of a critical
imaginary electric field).

The triangular lattice case is particularly
interesting, as it proved to be exactly solvable by Baxter \BAXHH,
who explicitly worked out various critical exponents characterizing
the crystallization transition.  He found the two
critical points $z_\pm=({1\pm\sqrt{5}\over 2})^5$,
with respective thermal exponents $\alpha_-=7/6$ and $\alpha_+=1/3$ governing the singularities
of the thermodynamic  free energy $f\vert_{sing}\sim |z-z_i|^{2-\alpha_i}$,
$i=\pm$. These correspond respectively to the minimal CFT's with $(p,q)=(2,5)$ 
and $(p,q)=(5,6)$
respectively, namely with central charge $c(2,5)=-22/5$ and $c(5,6)=4/5$. 
These are the universality
classes of the 2D Yang-lee edge singularity \CARTWO\ 
and of the 2D critical three-state Potts model 
(a generalization of the Ising model with three spins instead of two).
 
As to the two other cases of hard squares and triangles, no exact solution is known 
to this day. The powerful
Corner Transfer Matrix method of Baxter has allowed for many exact series expansions
for thermodynamic quantities up to quite large orders \BAXHH\ \BAX\ eventually leading to 
the very reasonable conjecture that in these two bipartite cases the crystallization
transition point lies in the universality class of the critical Ising model, at
some critical value $z_+$ depending on the lattice, while the other critical
point still lies in the Yang-Lee edge singularity class, at some $z_-$ also depending
on the lattice.

Beyond numerical evidence accumulated so far, we will propose here a ``gravitational"
proof of the identification of these universality classes via their random
lattice version.

\subsec{Matrix models as combinatorial tools}

Matrix models are remarkably efficient tools to realize the summation over
(possibly decorated) tessellations of given genus 
of the type of \twodqgpf\ (see e.g. \MMC\ and \EY\ for
reviews). The basic idea is to replace the task of performing
say a Gaussian integration over an $N\times N$ Hermitian matrix $M$ by the
drawing of pictures, called ``Feynmann diagrams", weighted accordingly
in such a way that the weighted sum over diagrams equals the Gaussian
integral. 

\fig{The pictorial representation for matrix elements $M_{ij}$ (double half-edges
carrying matrix indices),
terms Tr$(M^4)$ (a collection of four double half-edges connected at a vertex,
with matrix indices conserved along the oriented lines), and the propagators 
used to glue pairs of double half-edges into edges in the fatgraphs representing
the matrix integral.}{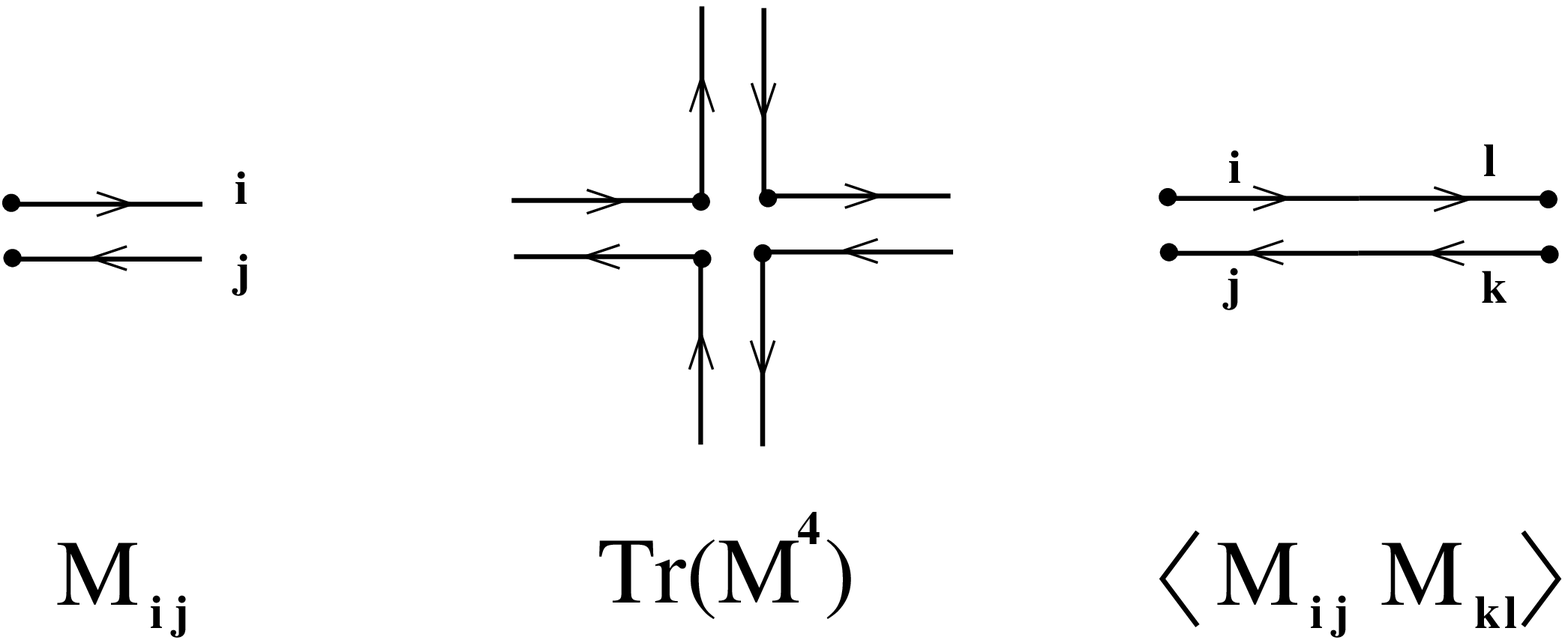}{9.cm}
\figlabel\reps

As an archetypical example of such a thing, let us display the quartic one-matrix
model, with partition function
\eqn\quartex{ Z(g)=\int dM e^{-N{\rm Tr}({M^2\over 2} -g {M^4\over 4})} }
where $dM$ stands for the standard Haar measure, normalized in such a way that
$Z(g=0)=1$. $Z(g)$ is to be understood as a formal power series of $g$,
the coefficients of which must be computed by performing Gaussian matrix integrals. 
By virtue of Wick's theorem, this is readily done (for the coefficient $Z_n$ of $g^n$)
by computing
\eqn\coefqua{\eqalign{
Z_n&= \int dM e^{-N{\rm Tr}({M^2\over 2})} {N^n\over n!} {\rm Tr}({M^4\over 4})^n\cr 
&= {(N/4)^n\over n!}
\sum_{pairs\ (ij),(kl)} \prod_{ij,kl} \langle M_{ij} M_{kl} \rangle \cr}}
where the sum extends over all decompositions of the matrix elements into pairs
and the notation $\langle M_{ij}M_{kl}\rangle={1\over N}\delta_{il}\delta_{j,k}$
stands for the ``propagators" $\int dM e^{-N{\rm Tr}(M^2/2)} M_{ij}M_{kl}$.
In turn, the value of the propagator allows to devise a pictorial representation
for the sum in \coefqua. Representing a matrix element $M_{ij}$ by a double
half-edge as in Fig. \reps, where each oriented line carries a matrix index
running from $1$ to $N$, we may express Tr$(M^4)$ as in Fig. \reps,
and interpret the sum in \coefqua\ as that over all possibly disconnected fatgraphs
obtained by gluing half-edges into edges (propagators). Each such graph
receives a weight $1/N$ per edge (from the propagators), $N g$ per vertex,
and $N$ per remaining oriented loop carrying an index summed from $1$ to $N$.
This gives an overall weight $g^A$ $N^{F-E+A}$ for each possibly disconnected graph
($A=\#$vertices, $F=\#$faces and $E=\#$edges),
as well as an overall rational factor expressed as $\sum 1/(4^n n!)$, which is
nothing but the inverse of the order of the automorphism group of the corresponding
fatgraph. Finally upon taking the logarithm of $Z(g)$, one ends up with the desired
sum over {\it connected} tessellations $\Theta$ (made of squares here, dual to the tetravalent
vertices of the matrix model), and with a weight $N^{\chi(\Theta)}$ $g^{A(\Theta)}$
$/|Aut(\Theta)|$. This particular example therefore corresponds to pure gravity,
in the form of random quadrangulations. Note that the size $N$ of the matrix 
serves as Newton constant in this approach, while the cosmological constant $g$
is a feature of the potential (term in the non-gaussian part of the exponential weight). 

More generally, we may engineer matrix models to suit our needs, either by considering
a more general potential allowing for weighted vertices of arbitrary valencies,
or/and by considering multi-matrix integrals of the form
\eqn\multimat{ Z=\int dA_1 ...dA_k e^{-N{\rm Tr}({1\over 2} \vec{A}^t Q \vec{A} 
-\sum_{i=1}^k V_i(A_i))} }
where $\vec{A}$ denotes the column vector of matrices $(A_1,...,A_k)^t$ and
$Q$ is a $k\times k$ quadratic form (with
$\vec{A}^t Q \vec{A}=\sum_{1\leq a,b\leq k} Q_{ab}A_aA_b$), 
and the $V_i$ some polynomial potentials. 
Note that again we normalize the measure in such a way that $Z=1$ if the $V_i$ vanish.
The diagrammatic interpretation of $Z$ follows from the multi-matrix generalization
of Wick's theorem expressing the multi-Gaussian matrix integral as a sum over 
pair decompositions of matrix elements of the integrand evaluated by using
propagators, namely
\eqn\wickgen{ \langle f(\{(A_a)_{ij}\}) \rangle_Q =\sum_{pairings} \prod_{pairs}
\langle (A_a)_{ij} (A_b)_{kl} \rangle_Q }
where $\langle ... \rangle_Q=\int \prod_a dA_a  ... e^{-N{\rm Tr}({1\over 2} 
\vec{A}^t Q \vec{A})}$ 
stands for the multi-Gaussian average. In particular, propagators are easily computed
in terms of the {\it inverse} of the quadratic form $Q$:
\eqn\propagen{ \langle (A_a)_{ij} (A_b)_{kl} \rangle_Q={\delta_{il}\delta_{jk}\over N}
(Q^{-1})_{a,b} }
The diagrammatic interpretation of $Z$ is now clear: attach different ``colors"
$a=1,2,...,k$ to the matrices. The potentials $V_i(x)=\sum g_{i,m} x^m/m$ contain 
the list of allowed vertices of color $i$ and valence $m$ with their respective weights
$g_{i,m}$. The quadratic form $Q$ encodes via its inverse the allowed propagators,
namely the list of possible edges linking these vertices to one another. The result
is a sum over vertex-colored graphs with specific gluing rules. 
Note moreover that irrespectively of its color each vertex receives a weight $N$, 
each edge a weight $1/N$ and each loop of oriented index lines a factor $N$, which 
still produces the topological factor $N^\chi$ in the expansion of
Log$\, Z$, $\chi$ the Euler characteristic of the (decorated) connected graph.
This allows in particular to extract the contribution from planar graphs
by taking $N\to \infty$.

In the next two sections, we use some of these models to solve the hard-particle
problem on various random lattices.

\subsec{Hard objects on random tetravalent graphs: matrix model solution}

\fig{A typical configuration of hard particles on a tetravalent planar graph.
The particles are represented by filled circles ($\bullet$). They obey the 
nearest neighbor exclusion constraint that no two particles may be adjacent
to the same edge.}{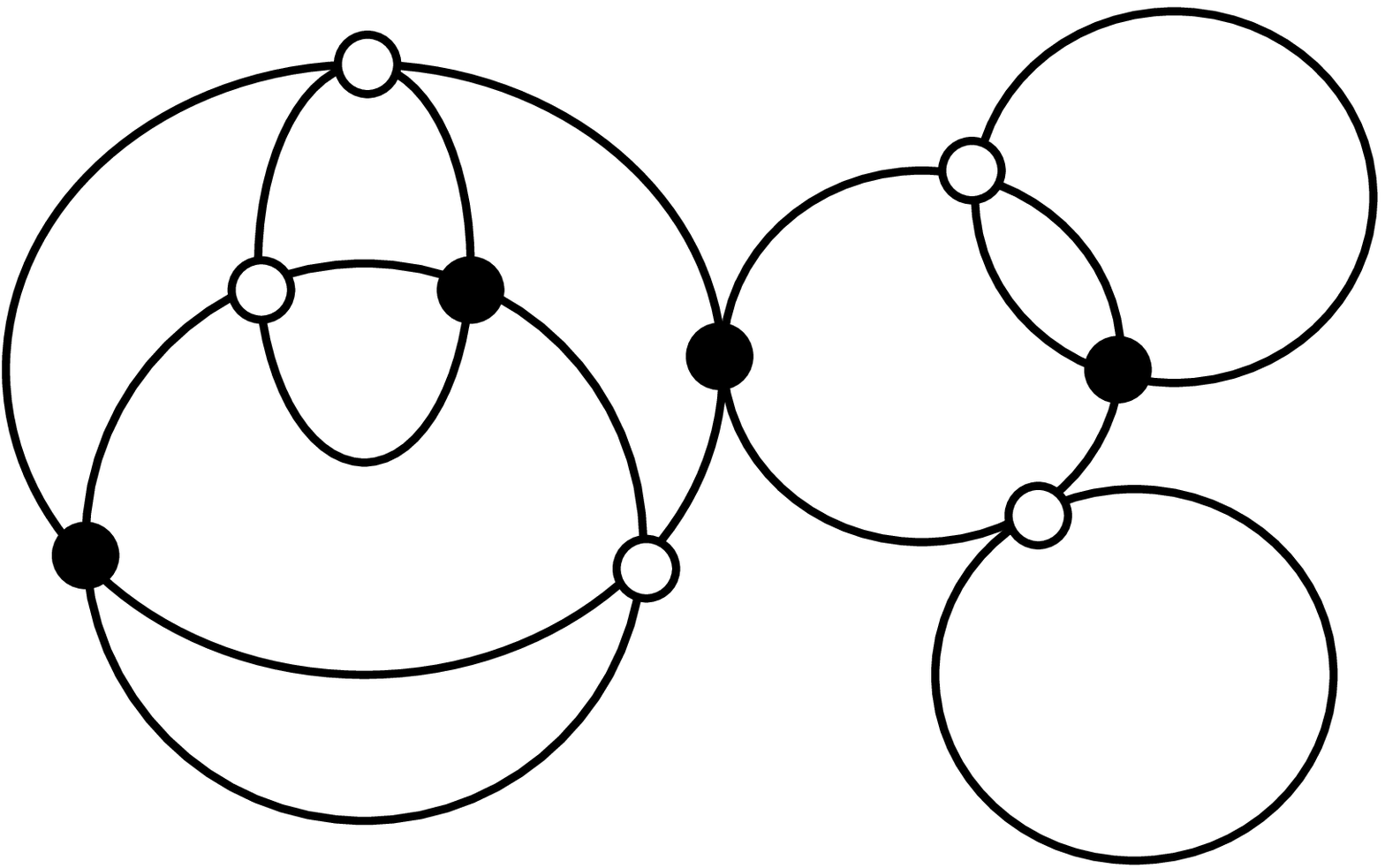}{8.cm}
\figlabel\hardconf

We now turn to the enumeration of hard-particle configurations on random tetravalent
graphs, the natural fluctuating versions of the square lattice. For illustration
we display in Fig. \hardconf\ a typical hard-particle configuration on a connected planar
tetravalent graph. The occupied vertices (represented by filled circles) obey the
nearest-neighbor exclusion (hardness) constraint that no two particles may be adjacent
to the same edge. As explained before, we attach a weight $z$ per particle and
$g$ per vertex. 

\fig{The pictorial representation for the matrix elements of the two-matrix
model describing hard-particles on tetravalent fatgraphs. The occupied (resp. empty)
vertices correspond to the confluence of four $A$ (resp. $B$) matrix elements,
represented as solid (resp. dashed) double half-lines. We have also represented
the only two non-vanishing propagators allowing for gluing double
half-edges into edges of the final graph. These obey the particle exclusion
rule that no two adjacent vertices of the final graph may be simultaneously
occupied. This information is encoded in the quadratic piece of the
potential of the two-matrix model.}{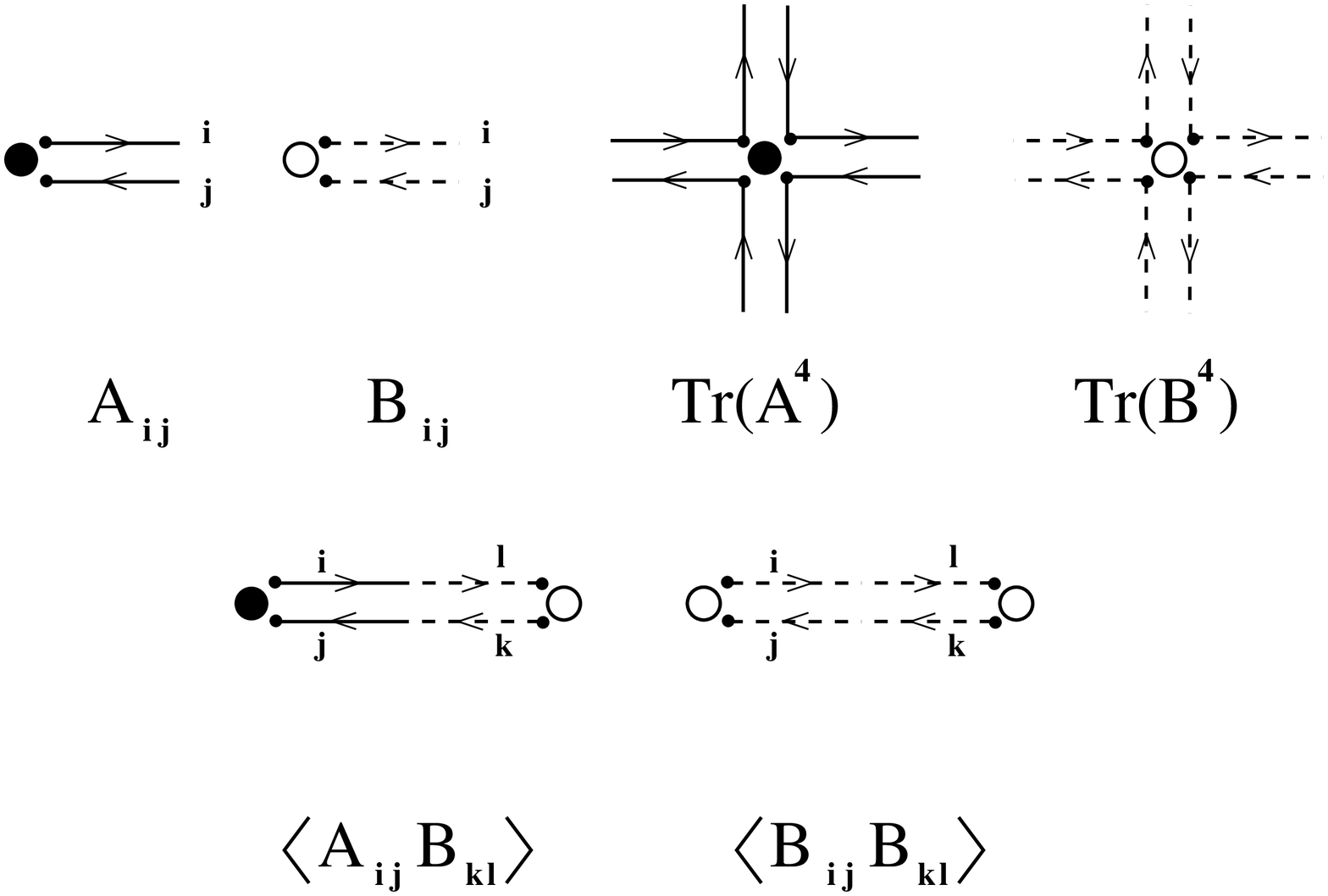}{10.cm}
\figlabel\hard

In view of previous section,
we are now ready to write a {\it two-matrix} model for enumerating such configurations, 
with partition function
\eqn\parthar{ Z(g,z)=\int dA dB e^{-N{\rm Tr}(AB-{A^2\over 2}
-gz {A^4\over 4}-g {B^4\over 4})} }
where the measure is now normalized in such a way that $Z(g=0,z)=1$. The graphical
interpretation (see Fig. \hard)
is a consequence of the two-matrix form of the Wick theorem,
in which propagators are calculated using the Gaussian weight 
$e^{-N {\rm Tr}(AB-A^2/2)}$, namely with $\langle A_{ij} B_{kl}\rangle =\delta_{il}\delta_{jk}/N$
and $\langle B_{ij} B_{kl}\rangle =\delta_{il}\delta_{jk}/N$ while 
$\langle A_{ij} A_{kl}\rangle =0$ as a particular case of eqn. \propagen. 
Representing the terms
Tr$(B^4)$ as empty tetravalent vertices and Tr$(A^4)$ as occupied ones, 
we see that the gluing of
half-edges
into edges (through the propagators) now incorporates the hard-particle exclusion rule
that no two occupied vertices may be adjacent (vanishing of the $\langle AA\rangle$
propagator). Again, Log$\, Z(g,z)$ is the generating function
for connected fatgraphs with hard particles on them, and with a weight
$g$ per vertex, $z$ per occupied vertex, times 
$N^{\chi(\Gamma)}/|Aut(\Gamma)|$ for the decorated graph $\Gamma$. 

The calculation of the matrix integral \parthar\ goes through a succession of standard steps.

$\bullet$ {\bf Step 1} consists in replacing the integral by one over eigenvalues of the matrices,
using the Itzykson-Zuber formula \ITZ. One is left with
\eqn\rediz{ Z(g,z)=\int d^Na d^Nb \Delta(a)\Delta(b) e^{-N{\rm Tr}(ab-{a^2\over 2}
-gz {a^4\over 4}-g {b^4\over 4})} }
where $a$ and $b$ are real diagonal matrices of size $N\times N$, and $\Delta$ stands
for the Vandermonde determinant $\Delta(a)=\det [a_i^{j-1}]_{1\leq i,j \leq N}$. 

$\bullet$ {\bf Step 2} consists in first noting that 
$\Delta(a)=\det [p_{j-1}(a_i)]_{1\leq i,j \leq N}$,
for any sequence of monic polynomials $p_j$ of degree $j$. One then picks the
two sets $p_j$ and $q_m$ to be inserted in $\Delta(a)$ and $\Delta(b)$ respectively
in such a way that they form a bi-orthogonal set wrt the following bilinear form:
\eqn\bilin{ (f,g)=\int dx dy e^{-N(xy-x^2/2-gz x^4/4-gy^4/4)} f(x) g(y) }
formally defined by using the size-one matrix propagators. Writing
\eqn\ortog{ (p_j,q_m)= \delta_{j,m} h_j(g,z) }
one then computes 
\eqn\calcfin{ Z(g,z)= \prod_{i=0}^{N-1} {h_i(g,z)\over h_i(0,z)} }

$\bullet$ {\bf Step 3} consists in calculating the $h_j$'s. One introduces the operators
$Q_1$ and $Q_2$ of multiplication by $x$ and $y$, expressed respectively
on the basis $p_j(x)$ and $q_m(y)$. Introducing similarly $P_1=d/dx$
and $P_2=d/dy$, one obtains upon integration by parts the following relations
\eqn\relatpq{\eqalign{{P_1\over N}&= Q_2^\dagger -Q_1- gz Q_1^3 \cr
{P_2\over N}&= Q_1^\dagger -g Q_2^3 \cr}}
where the adjoint is obtained by letting the corresponding operator
act on the other polynomial basis. 
As a consequence of \relatpq, the $Q$ operators take the form
\eqn\actQ{\eqalign{
Q_1 p_n(x)&= p_{n+1}(x)+ r_n p_{n-1}(x)+s_np_{n-3}(x)\cr
Q_2 q_m(y)&= q_{m+1}(y)+ {\tilde r}_m q_{m-1}(y)+{\tilde s}_mq_{m-3}(y)\cr}}
Note the presence of only every other term, due to the $\IZ_2$
symmetry of the potential under $A\to -A,B\to -B$.
The fact that $Q_1$ and $Q_2$ have a finite range is generic in multimatrix
models with polynomial potentials. 
The equations \relatpq\ finally turn into recursion relations for the various
coefficients of \actQ, which upon elimination allow for
expressing $v_n=h_n/h_{n-1}$, and finally the free energy
\eqn\pfinal{ F(g,z)={1\over N^2} {\rm Log} \, Z(g,z)=
{1\over N^2} \sum_{i=0}^N (N-i) {\rm Log}\left({v_i(g,z)\over v_i(0,z)}\right)}

$\bullet$ As we will be only interested in the large $N$ limit, in which only
planar graphs are selected (with genus zero free energy $f_0(g,z)=\lim_{N\to\infty}
F(g,z)$), let us perform the {\bf Step 4}, namely the reduction of equations in the
large $N$ case. We make the standard assumption that all coefficients $r_n,h_n$, etc...
tend to smooth functions of $x=n/N$ for $n,N\to \infty$ while $x$ remains fixed.
This is certainly true in the vicinity of the Gaussian model, for small $g$.
We have the limits
\eqn\resca{v_n\to v(x),\quad r_n\to r(x), \quad s_n\to s(x)}
and similarly for the tilded quantities. In the same limit, the $Q$ operators
\actQ\ become simply 
\eqn\qbecom{\eqalign{ Q_1&\to \sigma +r \sigma^{-1}+ s \sigma^{-3}\cr 
Q_2&\to  \tau +{\tilde r} \tau^{-1}+ {\tilde s} \tau^{-3}\cr}}
where the commuting dummy variables $\sigma,\tau$ are inherited from the corresponding 
(non-commuting) index shift operator acting on the respective polynomial bases. 
In particular, the operator adjoint simply follows from $\tau^\dagger=v \sigma^{-1}$. 
Plugging \qbecom\ into \relatpq\ and noting that $P_1/N=x \sigma^{-1} +O(\sigma^{-2})$, 
we finally get a set of algebraic equations for the $r$'s and $s$'s as well
as their tilded counterparts. Defining rescaled versions $V=g v$, $R=g r$, $S=g^2 s$
and similarly for the tilded quantities, we get
\eqn\algeb{\eqalign{
gx&=V-R-3z(R^2+S)\cr
0&= {\tilde R}-V(1+3 z R)\cr
0&= {\tilde S}-z V^3\cr
0&=R-3 V {\tilde R}\cr
0&=S-V^3\cr}}
These algebraic equations were given a combinatorial interpretation in \HARDOB\ \CONCUR,
where the various functions $V,R,S,...$ were shown to generate decorated rooted trees.
After elimination, the system \algeb\ reduces to
\eqn\fineq{gx=\varphi(V)\equiv V(1-3zV^2)-{3 V^2 \over (1-9 z V^2)^2}}
For fixed $g$ and $z$ this equation defines upon inversion the unique function
$V(x)$ such that $V(x)=gx+O(g^2)$ at small $g$.
It encodes the asymptotic properties of the sequence $v_n$, hence
those of the $h_n$'s, and finally we have
\eqn\thremof{\eqalign{ f_0(g,z)=
\int_0^1 dx (1-x) {\rm Log}\left({V(x)\over g x}\right) \cr}}
where we have identified $v(0,z)=x$.

It is now a simple exercise to find the singularities of the thermodynamic
function $f_0(g,z)$. Noting indeed that
\eqn\dedder{ {d^2\over dg^2} g^3 {d \over dg} f_0(g,z)={g d \over dg} {\rm Log}
\left({V_{g,z}\over g} \right) }
where $V_{g,z}$ is simply the unique solution to $g=\varphi(V_{g,z})$
such that $V_{g,z}=g+O(g^2)$ (identical to $V(x=1)$), we see that  
the singularities of the free energy are those of $V_{g,z}$.  
Defining the critical cosmological constant by
\eqn\criti{\eqalign{ g_c(z)&=\varphi(V_c(z)) \cr
0&= \varphi'(V_c(z)) \cr}}
where $V_c(z)$ is the smallest value of $V$ where a maximum of $\varphi$ is attained,
we immediately get the critical behavior $f_0\vert_{sing}\sim (g_c(z)-g)^{2-\gamma}$
where $\gamma=-1/2$, as this generically corresponds to a square root singularity 
of $V$ from $(g-g_c) \sim \varphi''(V_c)(V-V_c)^2/2$. 
Eqns. \criti\ are easily parametrized as 
\eqn\parameq{ z={12 u(1+3u)^2\over (1-3 u)^8} \qquad g_c(z)={(1-3u)^4(1+10u-15u^2)\over
12 (1+3u)^2} }
This gives a critical curve ending at the point 
\eqn\twosol{\eqalign{
z_-&=-{25\over 8192}(11\sqrt{5}+25) <0 \cr
g_-&=g_c(z_-)={64\over 45}(13\sqrt{5}-29) \cr
V_-&=V_c(z_-)={32\over 75}(7\sqrt{5}-15)\cr}}
and where in addition we have $\varphi''(V_-)=0$, which implies a susceptibility
exponent $\gamma=-1/3$, as we now have an equation of the form
$g-g_-\sim \varphi'''(V_-)(V-V_-)^3/6$.

To conclude, we find only one critical value of $z$ at a negative value $z_-$,
corresponding to a non-unitary critical point.
The value $\gamma=-1/3$ can be carefully traced back to the coupling of
the Yang-Lee edge singularity with $c(2,5)=-22/5$ to 2DQG\foot{This is in apparent
contradiction with the formula \kpz\ which would lead to $\gamma_{str}=-3/2$.
This is due to the fact that the leading behavior observed in the matrix model solution 
corresponds to another {\it scale} set not by the identity operator (leading by
integration over the random surfaces to the area $A=\int Id d^2 z$ or to the
conjugated coupling $g$ via the off-critical term $(g_c-g) \int Id d^2 z$ in
the gravitational action) but by that with
the lowest (negative) conformal dimension, as is the case in all non-unitary theories
with $c(p,q)$, say with $q>p+1$. 
This operator has conformal dimension $h_0=(1-(p-q)^2)/(4pq)$, and gravitationally
dressed dimension $\Delta_0= (p-q+1)/(2p)$ 
from \kpzdim. It corresponds to a term $(\mu_c-\mu)\int \Phi_{h_0} d^2z$
in the gravitational action, where by a dimensional argument
we find $g_c-g \sim (\mu_c-\mu)^{1-\Delta_0}$. The exponent
$\gamma$ observed in the matrix model corresponds to the scaling in terms
of $\mu_c-\mu$. Once translated back into the
language of the correct cosmological constant $g_c-g$, we have
$f\vert_{sing}\sim (\mu_c-\mu)^{2-\gamma} \sim (g_c-g)^{2-\gamma\over 1-\Delta_0}=
(g_c-g)^{2-\gamma_{str}}$. This allows to finally identify $\gamma=-2/(p+q-1)$
for any minimal model, while $\gamma_{str}=(p-q)/p$ follows from \kpz. 
Here, for $\gamma=-1/3$ we must have $p+q=7$, hence $p=2,q=5$. 
The only other possibility would be
$p=3$ and $q=4$, but the corresponding model is unitary and could not have a negative
critical value of $z$ as in \twosol.}.
This shows that no crystallization transition takes place when we couple 
the hard-particle model to ordinary 2DQG.
 
Let us however complete the task of computing the free energy.
Using the Lagrange inversion formula, we may explicitly invert
\fineq\ and evaluate any function of $V_{g,z}$: 
\eqn\invfineq{ h(V_{g,z})=\sum_{n\geq 1}{g^{n}\over n} \oint {dw \over 2i\pi}
{h'(w)\over \varphi(w)^n} }
Expressing $f_0(g,z)=\sum_{n\geq 1} f_n(z) g^n$, and using eqn. \dedder, we find that 
$f_n(z)=\varphi_n(z)/((n+1)(n+2))$, where $\varphi_n(z)$ are generated by
Log$(V_{g,z}/g)=\sum_{n\geq 1} \varphi_n(z) g^n$. We may now use the Lagrange inversion
formula \invfineq\ for $h(V)=$Log$(V)$. This gives
\eqn\givphi{ \varphi_n(z) ={1\over n} \oint {dw \over 2i\pi w} {1\over \varphi(w)^n}}
This finally leads to the compact formula
for the free energy $f_n(z)$ for hard particles on planar tetravalent
fatgraphs with $n$ vertices, and a weight $z$ per particle:
\eqn\finresu{\eqalign{ f_n(z)&={1\over n(n+1)(n+2)} \sum_{l,p\geq 0 \atop
2(l+p)\leq n} {2n-2l-2p-1\choose n-1}\times \cr
&\times {2n-l-2p-1\choose l}{4n-2l-p-2\choose p} 3^{n-l}
z^{l+p}   }} 
 
As a side remark, we recover the result for pure tetravalent graphs when $z=0$,
namely $f_n(z=0)=3^n {2n \choose n}/(2n(n+1)(n+2))=N_n/n$, $N_n$ as in \quadA\ (the
discrepancy is due to the extra rooting in \quadA\ which implies $N_n=n f_n$).
More interestingly, if we take $z\to \infty$ in \finresu, we should tend to a
crystalline groundstate in which vertices of the graphs are maximally occupied. 
We find that only $f_{2n}$ survive, with the result
\eqn\bicolf{ f_{2n}= {3^{n-1}\over n(2n+1)(2n+2)} {3n \choose n} }
The number $(2n)f_{2n}$ counts the rooted bipartite tetravalent planar graphs
\TUT\ \SCH. 
Another way of understanding this result, is by looking directly at the matrix integral
\parthar. Indeed, performing the change of matrix variables $A\to A/z^{1\over 8}$,
$B\to B z^{1\over 8}$ and redefining $g\to g/\sqrt{z}$, we immediately see that
in the limit $z\to \infty$ the matrix integral \parthar\ reduces to
\eqn\redparthar{ Z(g,\infty)= \int dA dB e^{-N{\rm Tr}(AB -{g\over 4}(A^4+B^4))} }
which clearly enumerates the bipartite tetravalent graphs.

The lesson to be drawn from this section is that summing over {\it arbitrary}
tetravalent graphs destroys the crystallization transition of the
hard-particle model. However, the crystalline groundstates which would make this
transition possible dominate the sum at large $z$, and involve only {\it bipartite}
tetravalent graphs. 
The problem is that these graphs have a negligible contribution at all finite values of $z$.
This suggests to reduce the range of summation to
bipartite graphs from the very beginning, which is the subject of next section.

\subsec{Hard objects on bipartite trivalent graphs: matrix model solution}

Simply for pedagogical purposes, we will address here the problem of enumeration of
hard-particle configurations on {\it trivalent} {\it bipartite} fatgraphs.
The case of tetravalent graphs was solved in \HARD, and presents only technical
complications, whereas the final results concerning the crystallization transition and its
universality class are the same. So we choose to concentrate here on the case
of hard-triangles coupled to 2DQG.

We wish to generate bipartite trivalent graphs say with alternating black and white
vertices,
which in addition may be either occupied or empty.
To generate the desired decorated graphs, we now need four matrices $A_1,A_2,A_3,A_4$
each standing for half-edges connected to empty white, occupied black, occupied white and
empty black vertices.
The corresponding matrix integral reads
\eqn\htmodfour{\eqalign{
Z(g,z)&=\int dA_1 dA_2 dA_3 dA_4 e^{-N{\rm Tr}\, V(A_1,A_2,A_3,A_4)} \cr
V(A_1,A_2,A_3,A_4)&=A_1 A_2 -A_2A_3 +A_3 A_4 -g({A_1^3\over 3}+{A_4^3\over 3})
 -gz({A_2^3\over 3}+{A_3^3\over 3})\cr}}
The quadratic form in $V(A_1,A_2,A_3,A_4)$ has been engineered so as to reproduce
the correct propagators, namely that only black and white vertices are connected
in the Feynman diagrams ($\langle A_i A_j\rangle =0$ if $i$ and $j$ have the same parity)
and that two occupied vertices exclude one-another
($\langle A_2 A_3\rangle=0$).

We may now repeat the straightforward, though tedious, steps 1-4 of the previous section.

$\bullet$ {\bf Step 1} takes us to the eigenvalue integral
\eqn\eigenint{ Z(g,z)=\int d^Na_1 d^Na_2 d^Na_3 d^Na_4
\Delta(a_1) \Delta(a_4) e^{-N{\rm Tr}\, 
V(a_1,a_2,a_3,a_4)} }
where the $a_i$ are real diagonal matrices of size $N\times N$. 

$\bullet$ {\bf Step 2} goes through as well, by simply replacing the bilinear form \bilin\ by
\eqn\newbilin{ (f,g)=\int dx_1 dx_2 dx_3 dx_4 e^{-NV(x_1,x_2,x_3,x_4)} f(x_1) g(x_4)}

$\bullet$ {\bf Step 3}: we define analogously operators 
$Q_i,P_i$, $i=1,2,3,4$, acting on the left 
orthogonal polynomial basis for $i=1,2$ and on the right for $i=3,4$,
and satisfying the following system, obtained by integration by parts:
\eqn\partgensys{\eqalign{
{P_1\over N}&= \partial_{x_1}V(Q_1,Q_2,Q_3,Q_4)= Q_2-gQ_1^2\cr
{P_2\over N}&=0=\partial_{x_2}V(Q_1,Q_2,Q_3,Q_4)=Q_1-Q_3-gz Q_2^2\cr
{P_3\over N}&=0=\partial_{x_3}V(Q_1,Q_2,Q_3,Q_4)=Q_4-Q_2-gz Q_3^2\cr  
{P_4\over N}&= \partial_{x_4}V(Q_1,Q_2,Q_3,Q_4)= Q_3-gQ_4^2\cr}}
The obvious symmetry of the potential under the interchange $x_i\leftrightarrow x_{5-i}$
allows to infer that $Q_4=Q_1^\dagger$ and $Q_3=Q_2^\dagger$, so that the main equations
reduce simply to
\eqn\simrednew{\eqalign{{P_1\over N}&= Q_2 -g Q_1^2 \cr
Q_1&= Q_2^\dagger +g z Q_2^2 \cr}}
The corresponding $Q$-operators are easily expressed on the left basis of orthogonal
polynomials as
\eqn\ragepq{\eqalign{
Q_1&= \sigma+ \sigma^{-2}r^{(1)}+\sigma^{-5}r^{(2)}+\sigma^{-8}r^{(3)} \cr
Q_2&=  \sigma^2 s^{(0)}+\sigma^{-1} s^{(1)}+\sigma^{-4}s^{(2)} \cr}}
where $\sigma$ is the shift operator acting on the $p$'s as $\sigma p_n=p_{n+1}$
and the operators $r^{(i)},s^{(i)}$ are diagonal. The presence of only powers of $\sigma$ spaced
by multiples of $3$ is a consequence of the accidental $\IZ_3$-symmetry of the potential $V$, 
namely thet $V(\omega x_1,{\bar \omega} x_2,\omega x_3,{\bar \omega} x_4)=V(x_1,x_2,x_3,x_4)$
for $\omega=e^{2i\pi \over 3}$.
The equations \simrednew\ together with the fact that $P_1/N\sim
(n/N)\sigma^{-1}+O(\sigma^{-2})$ allow in principle for solving the model for all $N$
by expressing recursion relations for the quantity $v_n=h_n/h_{n-1}$.

$\bullet$ {\bf Step 4}:
We are however only interested here in the planar $N\to \infty$ limit, in which
$\sigma$ becomes a commuting dummy variable, and the $r^{(i)},s^{(i)}$ some
functions of $x=n/N$, while $n,N\to \infty$. 
The equations \simrednew\
translate simply into algebraic equations for these functions, as well as for 
$v(x)=\lim v_n$. Upon defining the rescaled quantities 
$R^{(1)}=g^3 z^2 r^{(1)}$,
$S^{(0)}=s^{(0)}/g$,
$S^{(1)}=g^2 zs^{(1)}$,
$S^{(2)}=g^5 z^2s^{(2)}$
and $V=g^2 z v$, we find the algebraic system
\eqn\algebsys{\eqalign{
S^{(0)}&=1\cr
S^{(2)}&=-z V^4\cr
V&= S^{(1)} +2 S^{(1)} V\cr
R^{(1)}&= V^2 +z (S^{(1)})^2 +2 z S^{(2)}\cr
g^2z^2 x&= z S^{(1)} -2 R^{(1)}\cr}}
By elimination, we are simply left with
\eqn\finmas{ g^2z^2 x= \varphi(V)\equiv z {V\over (1+2V)^2} -2 V^2(1-2 V^2) }
while the desired planar free energy reads $f_0(g,z)=\int_0^1 (1-x){\rm Log}(V/(g^2 z x))$
by virtue of the obvious generalization of \thremof. The function $V$ must be
computed by inverting \finmas\ order by order in $g^2 x$. As in the case of previous 
section, the singularities of the planar free energy come from those of the solution
$V_{g,z}$ of \finmas\ at $x=1$, as we still have a relation of the form 
$d^2/dg^2 g^3 d/dg f_0(g,z)= g d/dg {\rm Log}(V_{g,z}/(g^2 z))$. 

\fig{The critical curves $g_c^2(z)$ are represented in solid ($g_2^2$)
and dashed ($g_1^2$) lines. The higher critical endpoint $z_-=-2^9/5^5$ and 
intersection point $z_+=32$ respectively correspond to the Yang-Lee edge
singularity and to the critical Ising model both coupled to 2DQG.}{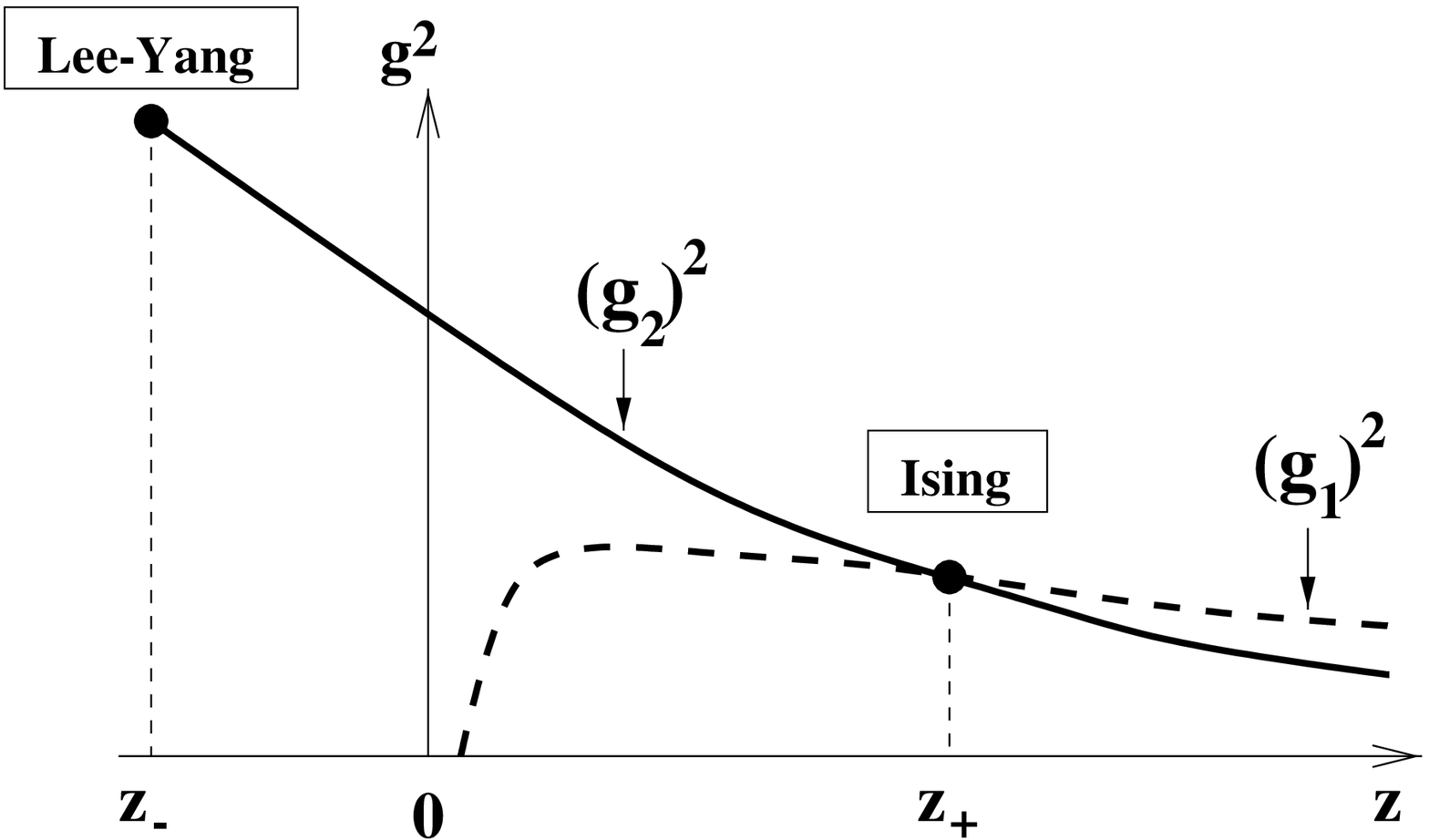}{8.cm}
\figlabel\twocurves

Expressing the critical points $g_c(z)$ solving the obvious adaptation
of \criti, namely that $g_c(z)^2=\varphi(V_c(z))/z^2$, while
$\varphi'(V_c(z))=0$, with the appropriate
value of $\varphi$ from \finmas, we see a first difference with the case
of previous section, namely that as
\eqn\phiprime{ \varphi'(V)\sim (1-2V)(z-4V(1+2V)^4) }
we now have {\it two} critical curves 
\eqn\twocrit{\eqalign{
g_1^2(z)&={1\over 8 z}-{1\over 4 z^2}, \ \ z>0, \ \ V=1/2\cr
g_2^2(z)&={1+8V+10 V^2\over 8(1+2V)^8}, \ {\rm with} \ z(V)=4V(1+2V)^4 \cr}}
along which we generically have a string susceptibility exponent
of $\gamma=-1/2$.
These are displayed in Fig. \twocurves. 
While the curve $g_2^2(z)$ possesses an endpoint at a value $z=z_-=-2^9/5^5$
(where $V_-=-1/10$, and $g_-^2=3.5^7/2^{20}$) at which $\varphi''$ vanishes
and  which qualitatively resembles
that found in the previous section, we now get another critical point at the intersection of the
two curves $g_1^2(z)$ and $g_2^2(z)$ at $z=z_+=2^5$ (with $V_+=1/2$ and $g_+^2=15/2^{12}$)
where we also have a vanishing of $\varphi''$. Both critical points correspond to
$\gamma=-1/3$, but while the first one is at a negative value of $z$
(hence expected to be described by a non-unitary CFT), the other one is at a positive
value of $z$. The only two minimal model
candidate CFT's for $\gamma=-1/3$ are such that $p+q=7$ (from the relation
$\gamma=-2/(p+q-1)$). This leaves us with only
the Yang-Lee edge singularity, non-unitary with $c(2,5)=-22/5$, and the
critical Ising model, unitary with $c(3,4)=1/2$. 
Noting that the second critical point found separates two phases of low and high occupancy,
we deduce from the above discussion that the
endpoint corresponds to the Yang-Lee model coupled to 2DQG, while the intersection
point corresponds to the crystallization transition, and is described 
by the critical Ising model coupled to 2DQG.
This assertion is further confirmed by actually computing critical exponents
corresponding to various operators (see \HARD\ for details).
This completes our gravitational proof that the hard triangle model's crystallization
transition lies in the universality class of the 
critical Ising model.

Let us for completeness express the planar free energy's coefficients $f_{2n}(z)$
of $g^{2n}$
in the (even) $g$-expansion, by inverting \finmas\ via the Lagrange inversion formula.
Repeating the arguments of the previous section,
we arrive at 
\eqn\exactenum{ f_{2n}(z)={2^n\over n(n+1)(n+2)} \sum_{0\leq 2p\leq j\leq n}
\left(-{1\over 2}\right)^p {2n-j-1 \choose n-1}{n-j \choose p}{4n-2j\choose j-2p} z^j}
Note this time the presence of signs in the polynomial expression for $f_{2n}(z)$. 
Note also that for $z=0$ we recover the number of bipartite trivalent planar graphs with $2n$
vertices
$N_{2n}=2n f_{2n}(z=0)=2^{n-1} {2n \choose n}/((n+1)(n+2))$ dual to the number of Eulerian
triangulations of area $2n$ \TUT\ \SCH.
The above results also translate into large $n$ asymptotics of $f_{2n}(z)$ 
via \largA, namely that
\eqn\larasy{\eqalign{
f_{2n}(z)&\sim {g_c(z)^{-2n} \over n^{7\over 2}} \qquad {\rm with} \left\{
\matrix{ g_c(z)=g_2(z) & z_-<z<z_+\cr g_c(z)=g_1(z) & z>z_+\cr} \right.\cr
f_{2n}(z_\pm)&\sim {g_c(z_\pm)^{-2n}\over n^{10\over 3}} \qquad {\rm with}\ 
\left\{ \matrix{g_c(z_-)&=\sqrt{15}.{5^3\over 2^{10}} \cr
g_c(z_+)&={\sqrt{15}\over 2^6}\ \ \ \  \cr} \right.\cr}}

\subsec{Partial conclusion}

Using matrix model techniques, we have been able to observe how
geometrical constraints (on the hard-particle models) have made bipartiteness
of the random lattices relevant. Indeed, imposing bipartiteness of the underlying
graphs from the very beginning proves to be instrumental in restoring the expected
crystallization transition.  
This allows to give a formal gravitational proof in the cases of hard triangles
(as well as hard squares) that
this transition is indeed in the universality class of the 2D critical Ising model. 
Some generalizations of the models shown here have been worked out in \HARD\
and concern particle models subject to a weaker exclusion constraint. These allow
in principle to visit all critical points described by minimal models
coupled to 2DQG, and must all be defined on bipartite graphs as well.  

One could wonder what happens in the case of hard hexagons. Inspired by the previous
lesson, we would expect that the crystallization transition for hard hexagons
only survives on {\it tripartite} (i.e. vertex-tricolored) fatgraphs.
We expect indeed the existence of three symmetric competing maximally occupied crystalline
groundstates to induce a transition in the universality class of the critical
3 state Potts
model coupled to 2DQG. Attempts in this direction show that a certain rectangular
matrix model should do the job of generating the corresponding configurations of
hard particles on tricolored trivalent graphs, but this model has not been amenable to
an exact solution yet \RECT.

As a both final and preliminary remark for the remainder of this note, the crucial
properties of {\it colorability} of the random lattices we have summed over have
guaranteed the existence of crystalline {\it groundstates}, also responsible
for the existence of the phase transition. In the next example of meanders, we will 
also arrive at a similar discussion of critical properties according to the
colorability constraints of the random lattices we will sum over. However
in this new case, not only the structure of groundstates does rely on
the colorability constraint, but the very definition of the degrees
of freedom of the model as well, essential in determining the central charge
of the underlying CFT.

\newsec{Meanders}

In this second part, we address the apparently unrelated problem
of enumeration of meanders, namely of topologically inequivalent configurations of
a road (simple closed curve) crossing a river (infinite line)
through a given number $2n$ of bridges (simple intersections).
After first describing the problem in detail, we will turn to a description
of the meander configurations as some particular random tetravalent planar
graphs decorated by loops (actually two loops: the road, and the river).
This leads us naturally to study the more general problem of loop models
coupled to 2DQG. In carefully pursuing this, we will eventually see that
geometrical constraints on these models (the so-called fully-packed loop models)
make again the {\it colorability} of the underlying lattice relevant, like
in the hard-particle case. We will be able to identify the
meander universality class by a subtle reasoning on the type of random lattices
we must sum over to finally generate the correct meander configurations.

\subsec{The meander enumeration problem}

Meanders are defined as planar graph configurations of a closed nonselfintersecting
curve (road) crossing a line (river) through a fixed number $2n$ of simple intersection
points (bridges). These configurations must be counted up to smooth deformations preserving the
topology. We denote by $M_{2n}$ the total number of such distinct configurations
for given number of bridges $2n$.
The meander enumeration is an old problem: it can probably be traced
back to  some work by Poincar\'e (1911), and reemerged in various contexts since:
as mathematical recreation \SL, as folding problem \TOU\ \DGG,
in relation to the 16th Hilbert problem \ARNO, in the theory of
invariants of 3-manifolds \KOSMO, in computer science \HMRT,
in abstract algebraic terms
\TLM\ \BACH, and in its own right \LZ\ \NOUS\ \GOL\ \MAK.

In the works \DGG\ and subsequent, our main motivation was the study of
the folding problem of polymer chains.
Such a polymer is ideally described by a chain of identical
line segments attached
by their ends, which serve as hinges between adjacent segments.
Think of a single strip of stamps, which can be folded along the
edges common to each neighboring stamps. We will distinguish between
closed and open polymers according to whether the chain forms a loop
or is open with two free ends. We will be addressing the
{\it compact self-avoiding} folding of such objects, namely
study the various ways in which the polymer can be completely
folded onto one of its segments.
Note that a closed compactly foldable polymer must have an
even number of segments.

\fig{Two compactly folded polymers and the corresponding meanders.
The first example is a closed polymer with $8$ segments, and corresponds
to a meander with  $8$ bridges.
The second example is an open polymer with $6$ segments,
and corresponds to a semi-meander with $7$ bridges.}{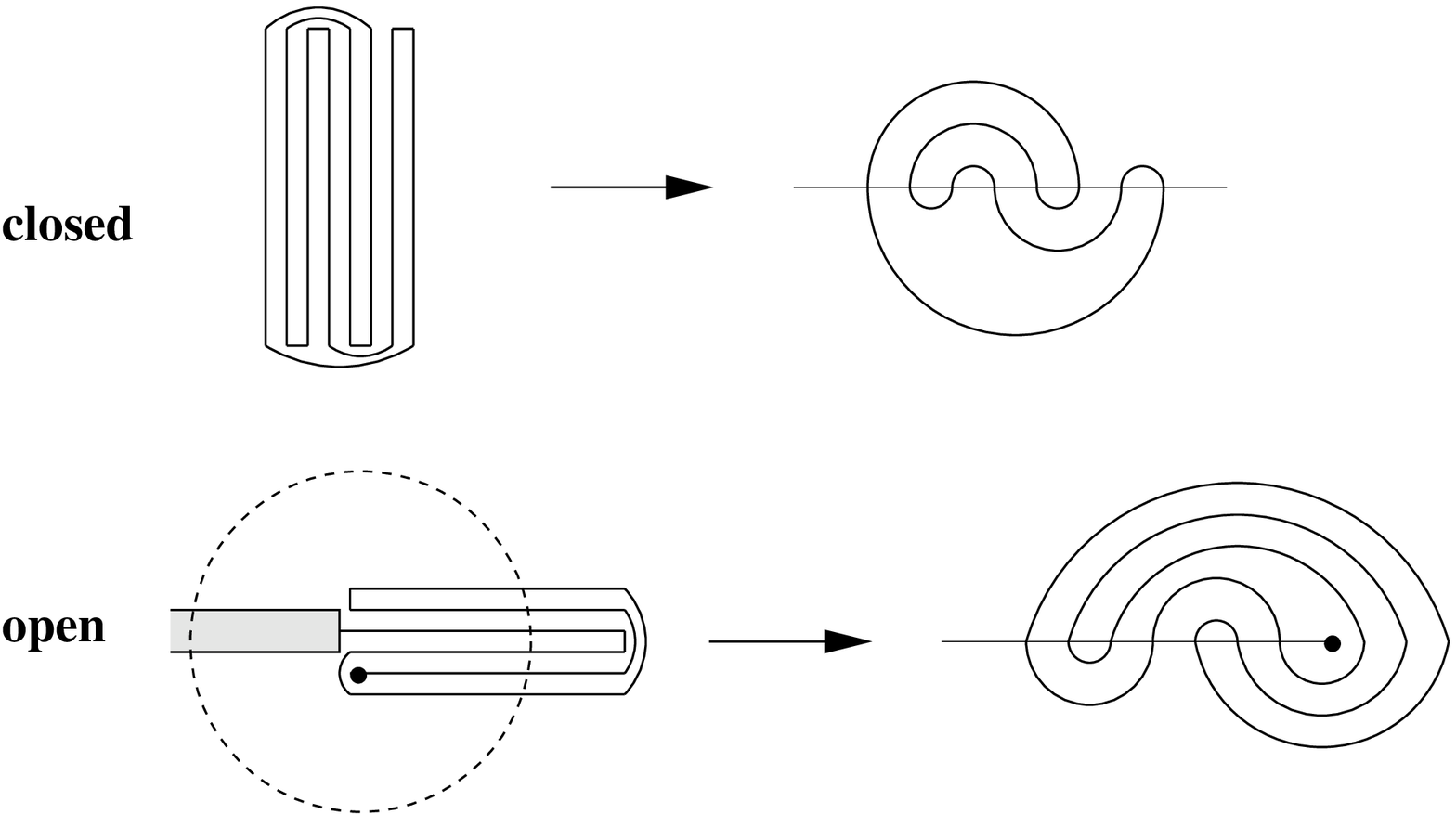}{8.cm}
\figlabel\separ

To distinguish between the various ways of compactly folding a
closed polymer, we will represent the folded objects as a
meander with $2n$ bridges.
To visualize the relation between compactly folded
closed polymers and meanders, it is simplest to imagine we draw a line
perpendicular to the segments forming the folded polymer with a total of
$2n$ intersections (each segment intersects the line once), and
then separate the various segments (see Fig.\separ).

In the case of an open polymer with say $n-1$ segments,
let us attach one of its ends to say a wall
or a support (see Fig.\separ),
so as to prevent the polymer from winding around that end (this is
exactly the situation  in a strip of stamps,
attached by one end to its support).
Starting from a compactly folded configuration, let us again draw
this time a circle that intersects each of the $n$ segments once, and also
intersects the support once. Extending the polymer so as to let it
form a half-line with origin its free end, we form
a planar configuration of a non-selfintersecting loop (road) crossing a
half-line (river with a source) through $n$ points.
These configurations considered up to smooth deformations preserving the topology
are called {\it semi-meanders} (see Fig.\separ\ for an illustration).
The total number of semi-meanders with $n$ bridges is denoted by ${\bar M}_n$.

The aim of the subsequent sections is to give a semi-rigorous physical argument
leading to the prediction of the meander and semi-meander configuration exponents
$\alpha,{\bar \alpha}$ which govern the large $n$ asymptotics of the
corresponding numbers $M_{2n}\sim g_c^{-2n}/n^\alpha$ and 
${\bar M}_n\sim g_c^{-n}/n^{\bar \alpha}$, $g_c$ some constant. This is done
by identifying the universality class of the corresponding critical phenomenon,
characteristic of the compact folding of one-dimensional objects.
We try here to summarize and put in perspective various works among which the paper
\ASY\ where these results were first obtained.

\subsec{Fixed lattice results for (fully-packed) loop models}

\fig{Qualitative phase diagram of the O(n) model on the honeycomb lattice.
The thick lines correspond to critical curves, for which we have indicated
the value of the central charge. We have parametrized $n=2\cos \pi e$,
$0\leq e\leq 1$.}{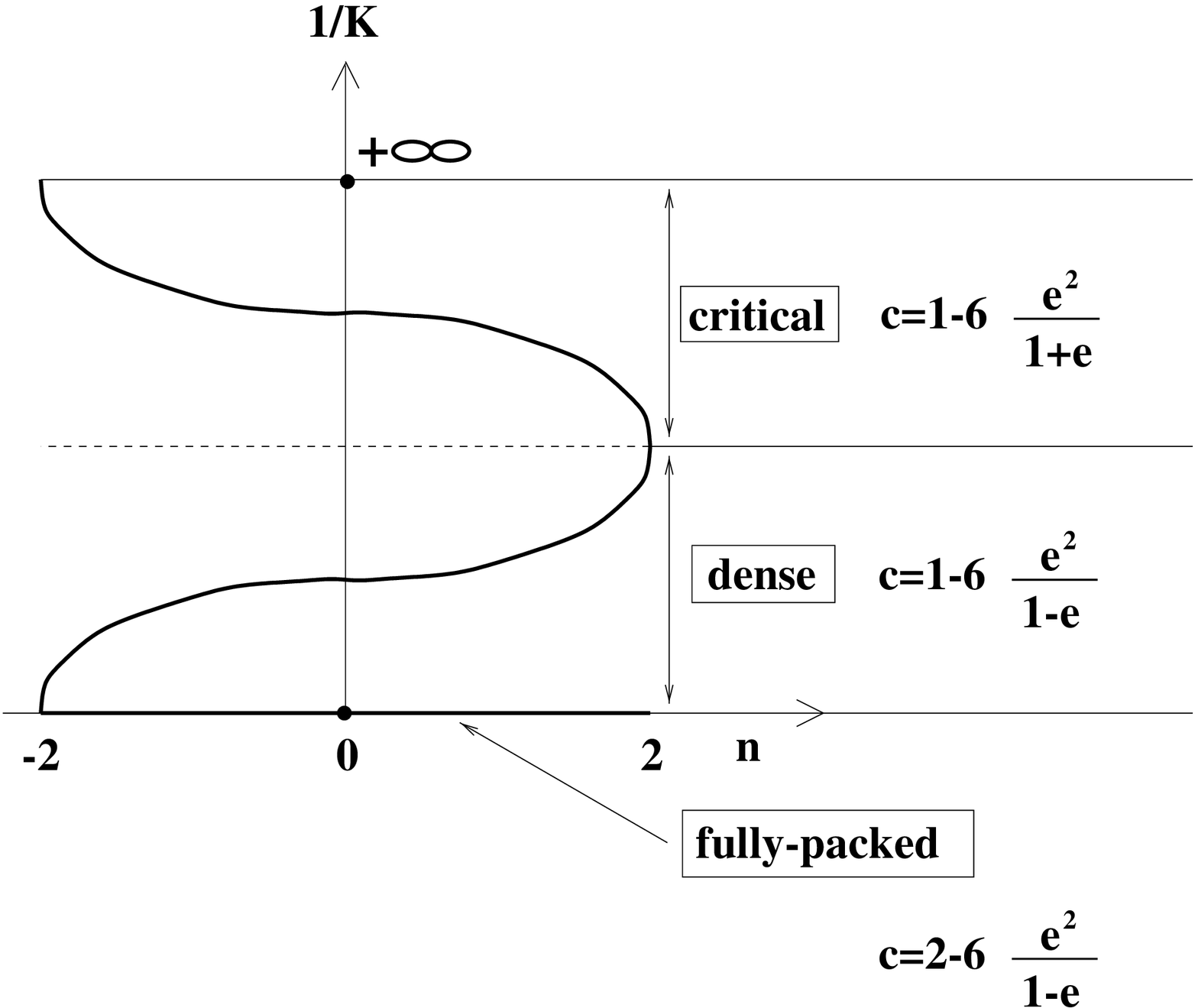}{10.cm}
\figlabel\onheycomb

Loop models have been studied extensively for decades as toy models
for describing (self-avoiding) polymers and we will only summarize
the results obtained for their critical behavior in two dimensions. 
They are defined as statistical lattice models, for which the configuration
maps $\sigma$ go from the set of edges of the lattice to ${\cal T}=\{0,1\}$,
where $\sigma(e)=1$ iff the edge $e$ is occupied by a loop edge. 

More specifically let us first discuss the case of loop (so-called O(n))
models on the honeycomb lattice. The self-avoidance constraint is very
easy to implement as at most one loop may visit any given vertex, hence at each
vertex with adjacent edges $e,e',e''$, we either have 
$\sigma(e)=\sigma(e')=\sigma(e'')=0$, or exactly two of the edges are occupied
say $\sigma(e)=\sigma(e')=1$, while $\sigma(e'')=0$. We attach two types
of Boltzmann weights to the loop configurations: a weight $K$ per occupied edge,
and a weight $n$ per loop. The partition function of the model reads
\eqn\parton{ Z(n,K)=\sum_{{\rm loop}\ {\rm configs.}} n^L \ K^E }
where $L$ is the total number of loops and $E$ the total number of loop edges.
The qualitative phase diagram of this model is displayed in Fig. \onheycomb.    
The model undergoes various phase transitions within the range $-2\leq n\leq 2$,
in which we parametrize $n=2\cos \pi e$, $0\leq e\leq 1$.
It has {\it three} critical curves (represented as a zig-zag thick line and a thick straight
line), characterized by the density of loops being either $0$, between $0$ and $1$
or $1$, respectively called dilute, dense and fully-packed phase. 
The latter case corresponds to a situation where the lattice is maximally occupied by loops,
namely each vertex is visited by a loop, and it is obtained in the limit $K\to \infty$.
The three critical
curves are respectively described by CFT's with respective central charges
\NIEN\ \BN\ \BSY\
\eqn\cchar{\eqalign{ {\rm dilute}: c&=1-6{e^2\over 1+e} \cr
{\rm dense}: c&=1-6{e^2\over 1-e} \cr
{\rm fully-packed}: c&=2-6{e^2\over 1-e} \cr}}
It is beyond the scope of these notes to explain in detail these three formulas.
They involve the so-called Coulomb gas description of the various critical theories,
namely a description involving scalar free fields, each contributing $1$ to the central
charge, while a suitably defined background electric charge accounts for the contribution
proportional to $e^2$. 
Typically, these scalar fields are the continuum limit of discrete ``height variables"
describing the degrees of freedom of the original model. In the case of loops, one
may indeed define a dual height variable as follows. 
We first orient the loops
arbitrarily, and then define a height variable 
as constant within each domain delimited by
loops, and increasing or decreasing by a fixed amount when one goes across a loop 
pointing to the left or right (in other words, the loops form
a contour plot of the height variable). 
This explains at least vaguely why only one scalar field
will in general be sufficient to give a complete description of loop models. The background
electric charge is also easy to understand as follows. To produce the correct weight
$n=2\cos\pi e$ per loop, one simply attaches a weight $e^{\pm i\pi e}$ for each 
orientation. 
In turn, this is obtained by simply introducing a local weight $e^{\pm i\pi e/6}$
on each vertex visited by a loop, according to whether the loop makes a left or right turn.
Indeed, on the hexagonal lattice the total of left minus right turns is always $\pm 6$,
and the weight $n$ per loop follows from the summation over all loop orientations.
Now if we try to make sense of the loop model in a cylindric geometry, we
see that if a given loop winds around the cylinder, it will have as many left and right
turns, hence receive a wrong weight $2$ instead of $n$. This is repaired by adding the
background electric charge, which modifies the free field action and restores the correct
weight, while modifying the central charge accordingly. 

The fully-packed case is clearly distinguished in that it requires
{\it two} scalar fields, rather than just one for the two other cases. 
More precisely,
the central charge for fully-packed loops is exactly {\it one more} than that of the dense
loops for the same value of $n$.
Let us show in the two particular cases $n=1,2$ where this extra degree of freedom comes from. 

\fig{Various interpretations of the fully-packed loop model's configurations on the 
hexagonal lattice: (i) rhombus tiling, obtained via the dual of the hexagonal lattice,
the triangular lattice, in which we erase the edges dual to unoccupied edges, thus forming
rhombi, which come in three possible orientations. (ii) Antiferromagnetic groundstates
of the Ising model: one tries to maximize the number of antiferromagnetic interactions
$(+-)$ between neighboring spins on the triangular lattice. Exactly one third of the 
edges remain frustrated (dashed here) namely are adjacent to equal spins: 
these frustrated edges
of the triangular lattice are nothing but the erased ones in the rhombus tiling.
(iii) Fully-packed loops on the hexagonal lattice: each vertex is visited
by exactly one loop.}{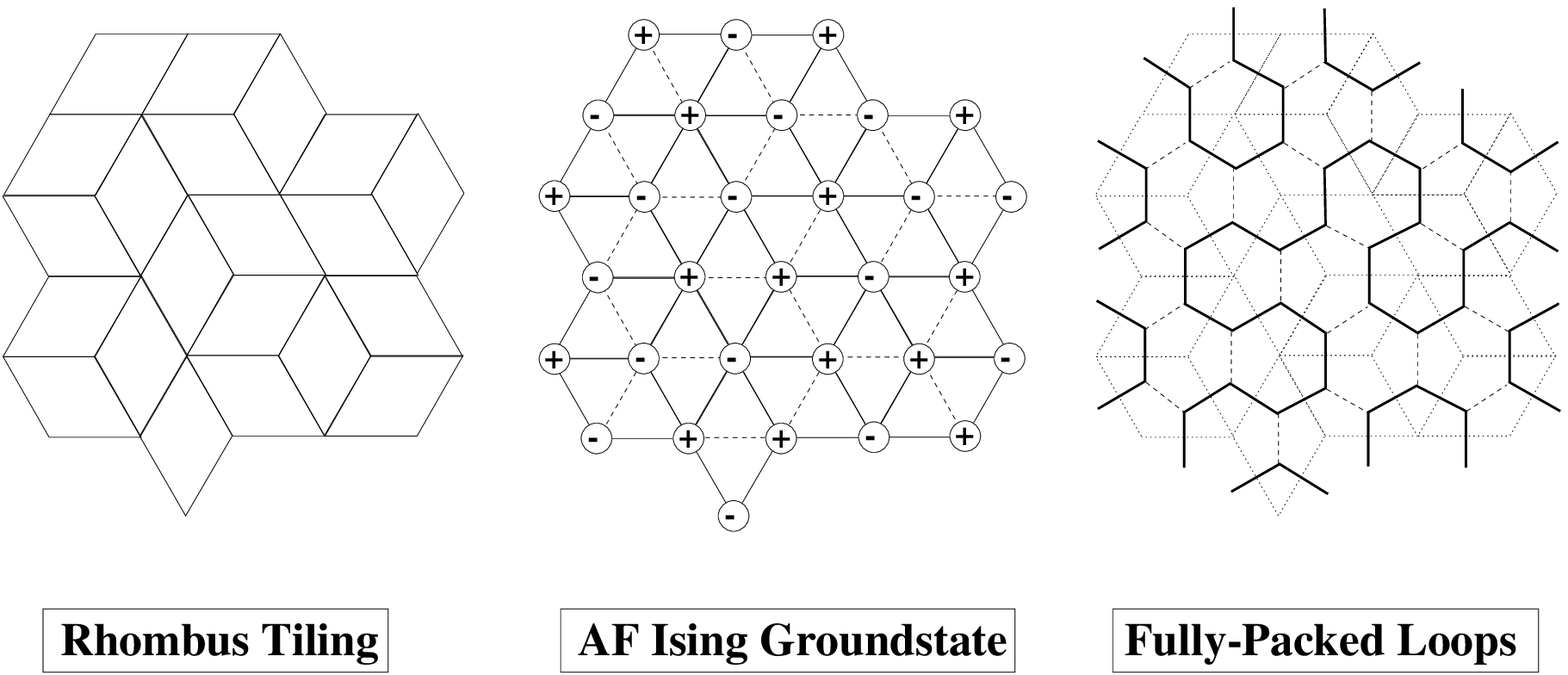}{13.cm}
\figlabel\onone

The case $n=1$ describes various equivalent problems for which
configurations are depicted in Fig. \onone: 
(i) rhombus tiling of the plane 
(ii) groundstates of the antiferromagnetic Ising model on the triangular lattice 
(iii) fully-packed loops on the honeycomb lattice.
The interpretation (i) allows to immediately understand why the central charge
$c=1$ in this case. Indeed, the rhombus tiling may also be regarded as a view
in perspective of the interface between two media, made of the piling-up of cubes.
The natural height variable for this problem is simply the actual height of the various
cubes, which in the continuum becomes the $z$ coordinate of some landscape. 

\fig{The triangular lattice is represented as a surface, with edges viewed as
tangent vectors (orientations are indicated by arrows), 
such that the sum over edges around each face vanishes 
$\sum_{\rm face} \vec{e}_i=\vec{0}$. A phantom
folding map $\sigma$ is simply a map sendig tangent vectors to either
of the three images $1,2,3$ that we have represented. This image is simply the 
value of the tangent vector in the folded configuration. 
As such, it must satisfy the face-rigidity constraint that $\sum_{\rm face}
\sigma(\vec{e}_i)=\vec{0}$.}{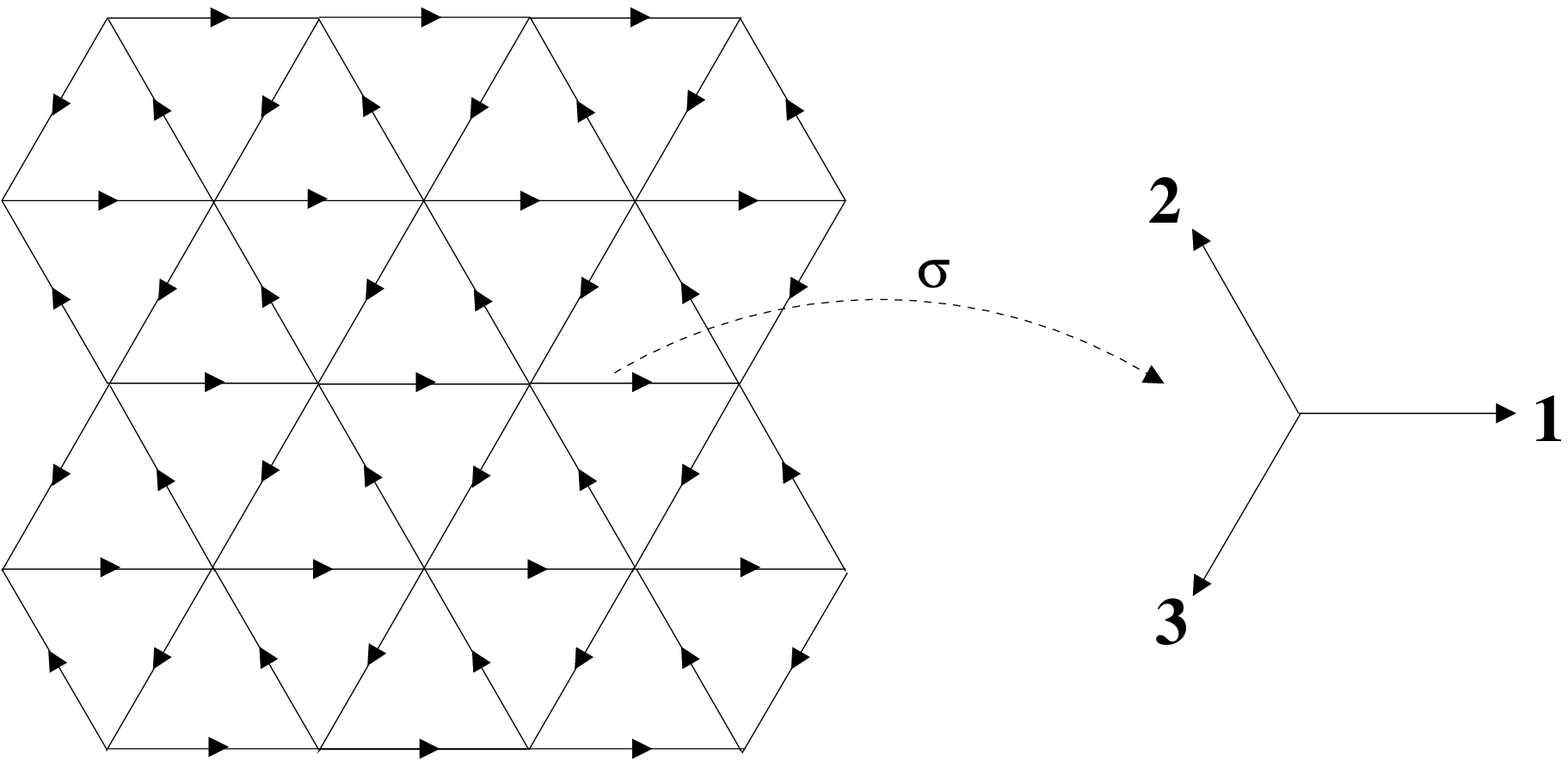}{9.cm}
\figlabel\tricol

The case $n=2$ may be interpreted as the problem of ``phantom folding" of the 2D triangular
lattice \GDF. We start from the triangular lattice, viewed as a surface, namely with edges drawn
as tangent vectors $\vec{e}$, in such a way that the sum of tangent vectors around each face
vanishes (see Fig. \tricol), namely
\eqn\folcons{\sum_{e\ {\rm around}\ {\rm face}\ f} \vec{e} =\vec{0} }
A phantom folding configuration is simply a configuration of this surface after being folded
back to the plane, in a process where faces are not deformed, and edges serve as hinges
between adjacent triangles. We are however interested only in the final folded states,
not in their realizability with an actual material such as a piece of paper. It means that
some of the foldings we consider may require cutting the sheet and letting it cross itself,
hence the name phantom. More precisely, a folding map $\sigma$
simply sends each edge $\vec{e}$ to its new
value in the folded configuration $\sigma(\vec{e})$. 
Finally the face-rigidity constraint turns into 
\eqn\facerig{ \sum_{e\ {\rm around}\ {\rm face}\ f} \sigma(\vec{e}) =\vec{0} }
It is easy to see that the images $\sigma(\vec{e})$
may only take one of three possible values, denoted
$1,2,3$ in Fig. \tricol, namely the three unit vectors with vanishing sum (up to a global
rotation). This allows to replace any phantom folding configuration 
by an edge tricoloring configuration of the triangular lattice, in which we paint
edges with colors $1,2,3$ in such a way that the three edges adjacent to each face 
have distinct colors.
Representing the dual of such a configuration, we obtain an edge-tricoloring of the
hexagonal lattice. Let us concentrate on the edges colored $1$ and $2$. It is easy to see
that edges of alternating colors $1,2,1,2,1,2...$ form fully-packed loops on the hexagonal
lattice. Now exchanging the colors $1\leftrightarrow 2$ along any of these loops
independently also produces an admissible and distinct edge-tricoloring. 
This shows that the fully-packed loops receive naturally a weight $n=2$ per loop.
Finally picking
arbitrarily the edges of color $3$ in such a way that there be exactly one per vertex
allows to generate all fully-packed loop configurations. We therefore end up with
the fully-packed loop model on the hexagonal lattice, with $n=2$.
The height variable of the folding model is easily identified as the actual
coordinate $\vec{h}$ (in $\IR^2$) of the vertices in the folded configurations.
Being two-dimensional, these give naturally rise to {\it two} component scalar fields
in the continuum, which explains the central charge $c=2$ in this case (no background charge
is necessary as there are no specific left/right turn weights).
Note that the edge-tricoloring problem of the hexagonal lattice was solved 
by Baxter \BAXTRI\ by Bethe Ansatz techniques, and the central charge $c=2$ may be
directly extracted from that solution.
The general fully-packed result \cchar\ was derived by analogous techniques in \BSY.

\fig{A typical Fully-Packed loop configuration on the square lattice. Assuming
doubly periodic boundary conditions, there are 6 black loops
(solid line) and 4 white ones (dashed lines). Up to rotations, the vertices of the
model are of the two types (a) ``crossing" or (b) ``tangent".}{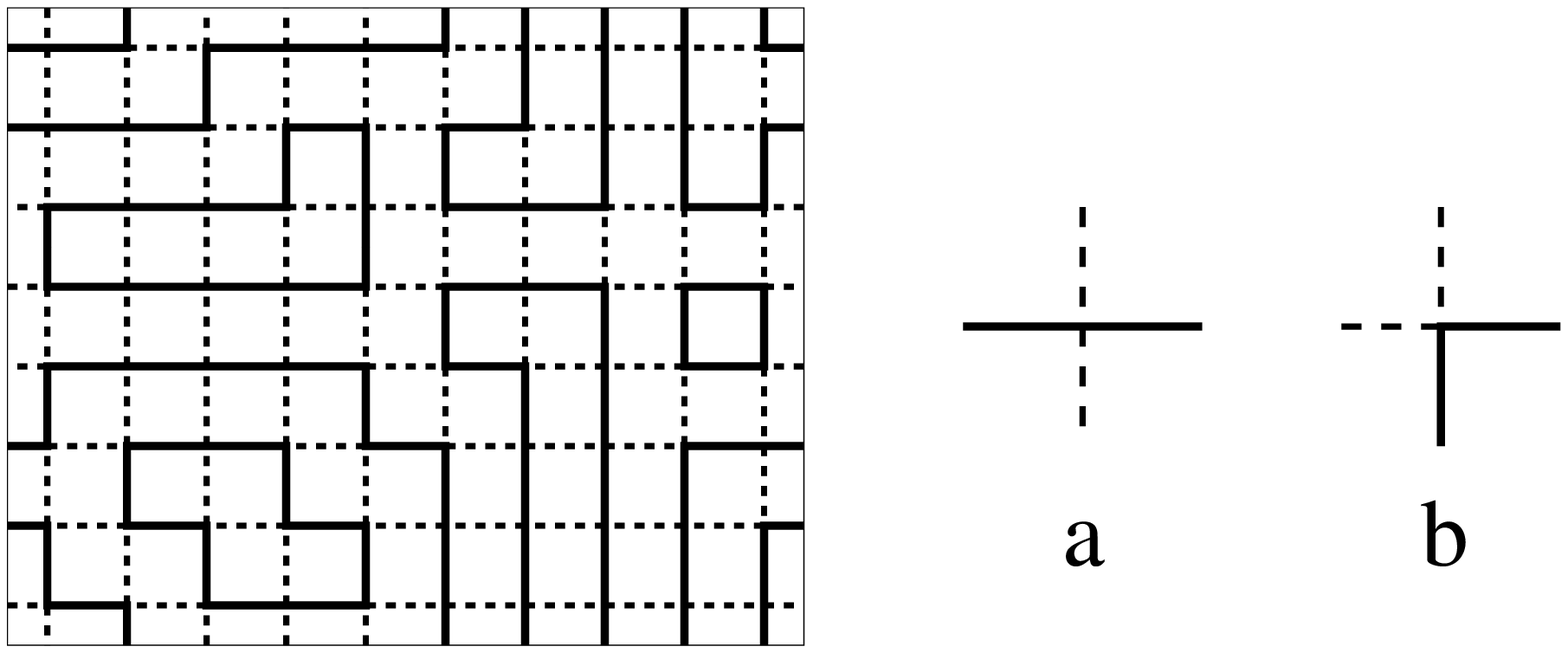}{9.cm}
\figlabel\vertex
 
Let us also mention another interesting loop model, which will eventually allow us to
solve the meander problem: the two-flavor loop model on the square lattice. The configurations
of this model are made of loops of {\it two} colors (flavors) say $1,2$ (or black,white) 
occupying edges of the square lattice, themselves belonging to at most one loop. 
The new condition is that although these loops are self-avoiding,
two loops of distinct colors may coexist at a given vertex, by either crossing one-another,
or being ``tangent to one-another" as displayed in Fig. \vertex\ (a) and (b). 
The model is further defined by attaching a weight $K_i$ per edge occupied by a loop
of color $i=1,2$ and $n_i$ per loop of color $i=1,2$.
The partition function of this model reads:
\eqn\partwofla{ Z(n_1,n_2;K_1,K_2)=\sum_{{\rm loop}\ {\rm configs.}} 
n_1^{L_1} n_2^{L_2} K_1^{E_1} K_2^{E_2} }
where $L_i$ is the total number of loops of color $i=1,2$ and $E_i$ the total number
of edges in the loops of color $i=1,2$.
In a way similar to the hexagonal lattice case, the model undergoes  
a number of interesting phase transitions, and in particular it still has
a dense critical line, as well as a fully-packed one, for $-2\leq n_1,n_2 \leq 2$, which
we still parametrize by $n_i=2\cos\pi e_i$, $0\leq e_i\leq 1$, $i=1,2$.
The fully-packed constraint restricts the model to only configurations where all
vertices are visited by the two types of loops, which corresponds to taking
$K_1,K_2\to \infty$ (see Fig. \vertex\ for an example).
This model is referred to as the FPL$^2$ model in the following.
As before, Coulomb gas techniques have been used to identify precisely the
universality classes of the various critical models, leading in particular
to formulas for the central charge of the underlying CFT \JACO:
\eqn\cenchartwo{\eqalign{
{\rm dense}: c&=2-6\left( {e_1^2\over 1-e_1}+{e_2^2\over 1-e_2}\right) \cr
{\rm fully-packed}: c&=3 -6\left( {e_1^2\over 1-e_1}+{e_2^2\over 1-e_2}\right) \cr}}
hence we still have an increase of $+1$ in the central charge when going 
from the dense to the fully-packed case at fixed $n_1,n_2$.
Again, the $2$ and $3$ in the central charges correspond to the number of degrees of freedom
necessitated by the field theoretical description of the continuum limit, while the 
other contributions come from background charges which ensure the correct weights
$n_i$ per loop of color $i$ in cylindric geometries. 
Let us now show where the three scalar fields come from in the fully-packed case.

\fig{A typical configuration of the FPL$^2$ model together with the
bicoloration of its vertices (checkerboard of filled ($\bullet$) and empty ($\circ$)
dots). We have added the corresponding dictionary that allows to map
the loop configurations onto $A,B,C,D$ labelings of the edges.}{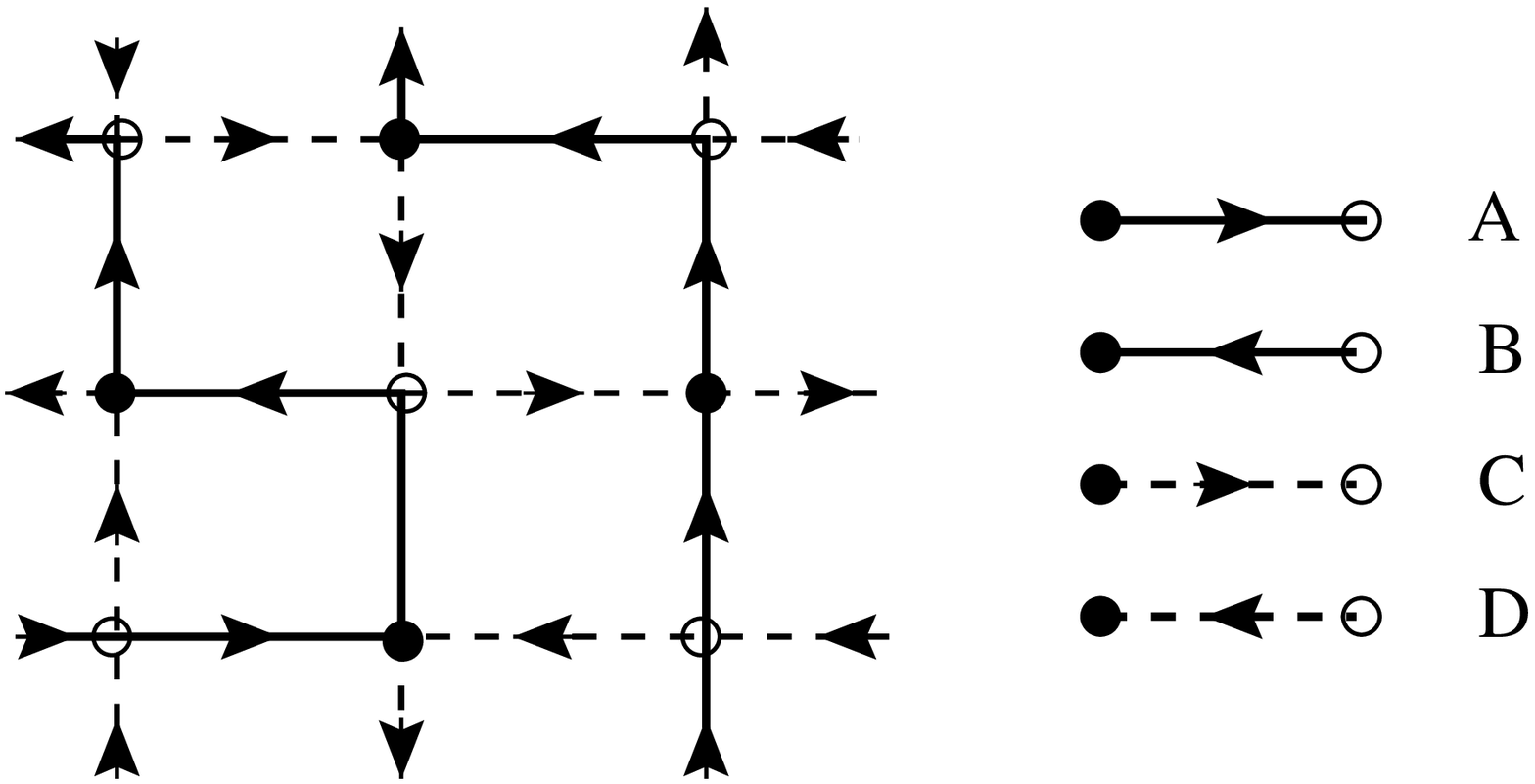}{7.cm}
\figlabel\dictio

To do this, let us rephrase the model as a (3D)
height model as follows.
Starting again from an {\it oriented} fully-packed black and white loop configuration,
we first {\it bicolor} the vertices
of the square lattice, say with alternating filled ($\bullet$) and empty ($\circ$) dots.
Then we use the dictionary of Fig.\dictio\
to assign one of the four labels $A,B,C,D$ to each colored and oriented edge.
With this convention, it is clear that edges of type $ABAB...$ alternate
along black loops, whereas
edges of type $CDCD...$ alternate along white loops, and that each vertex
has one incident edge of each type $A,B,C,D$.
It is clear that the four-labeling with $A,B,C,D$ is in
one-to-one
correspondence with the coloring {\it and} orientation of edges of the FPL model.
In particular, the orientation of a given black or white loop is reversed if we
interchange
the $A$ and $B$ or $C$ and $D$ labels along the loop.

\fig{Rules determining the change of the height variable across labeled edges.
We adopt the Amp\`ere convention that the height is increased (resp. decreased)
by the edge value if the arrow of the edge points to the left (resp. right).
The edge labels must be interpreted as three-dimensional vectors with the
respective values $\bf A$, $-\bf B$, $\bf C$, $-\bf D$}{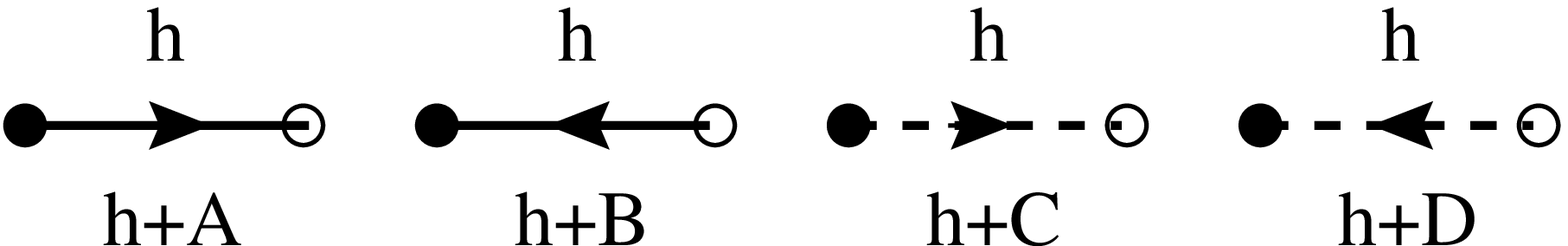}{9.cm}
\figlabel\ampere

The above colors allow for defining a dual vector height variable
on the center of each face of the lattice. Indeed, viewing as vectors the
$A,B,C,D$ labeling of the edges
of the lattice, let us arbitrarily fix the height to be zero
on a given face of the lattice, and define it on all faces by successive
use the rules of Fig.\ampere\
for the transition from a face to any of its neighbors.
Note that it is necessary to impose the condition ${\bf A}+{\bf B}
+{\bf C}+{\bf D}=0$ to
ensure that the heights are consistently defined around each vertex.
We may therefore assume in all generality that
${\bf A},{\bf B},{\bf C},{\bf D}$
are actually four vectors in $\IR^3$
with vanishing sum;
let us take for definiteness ${\bf A},{\bf B},{\bf C},{\bf D}$
to be the four unit vectors pointing from
the center of a tetrahedron towards its vertices.
The heights are then clearly three-dimensional, as linear combinations of
${\bf A},{\bf B},{\bf C},{\bf D}$.

To conclude this section, we have presented two types of loop models on the hexagonal
and square lattice and studied their critical lines. In both cases, the fully-packed phase
is described by {\it one more} scalar field than the dense one, resulting in
an increase of $+1$ in the CFT's central charge.

\subsec{Fully-packed loop models and 2DQG}

In the next section, we will show how meanders may be viewed as configurations of 
the two-flavor loop model on a random lattice. To pave the road for further results,
let us first consider here the ordinary (one-flavor) loop model on a random (trivalent)
lattice. 

\fig{Sample configurations of a planar non-Eulerian triangulation, and
a planar Eulerian one. The Eulerian triangulation has only vertices
of even valence, which allows to bicolor the triangular faces as shown.}{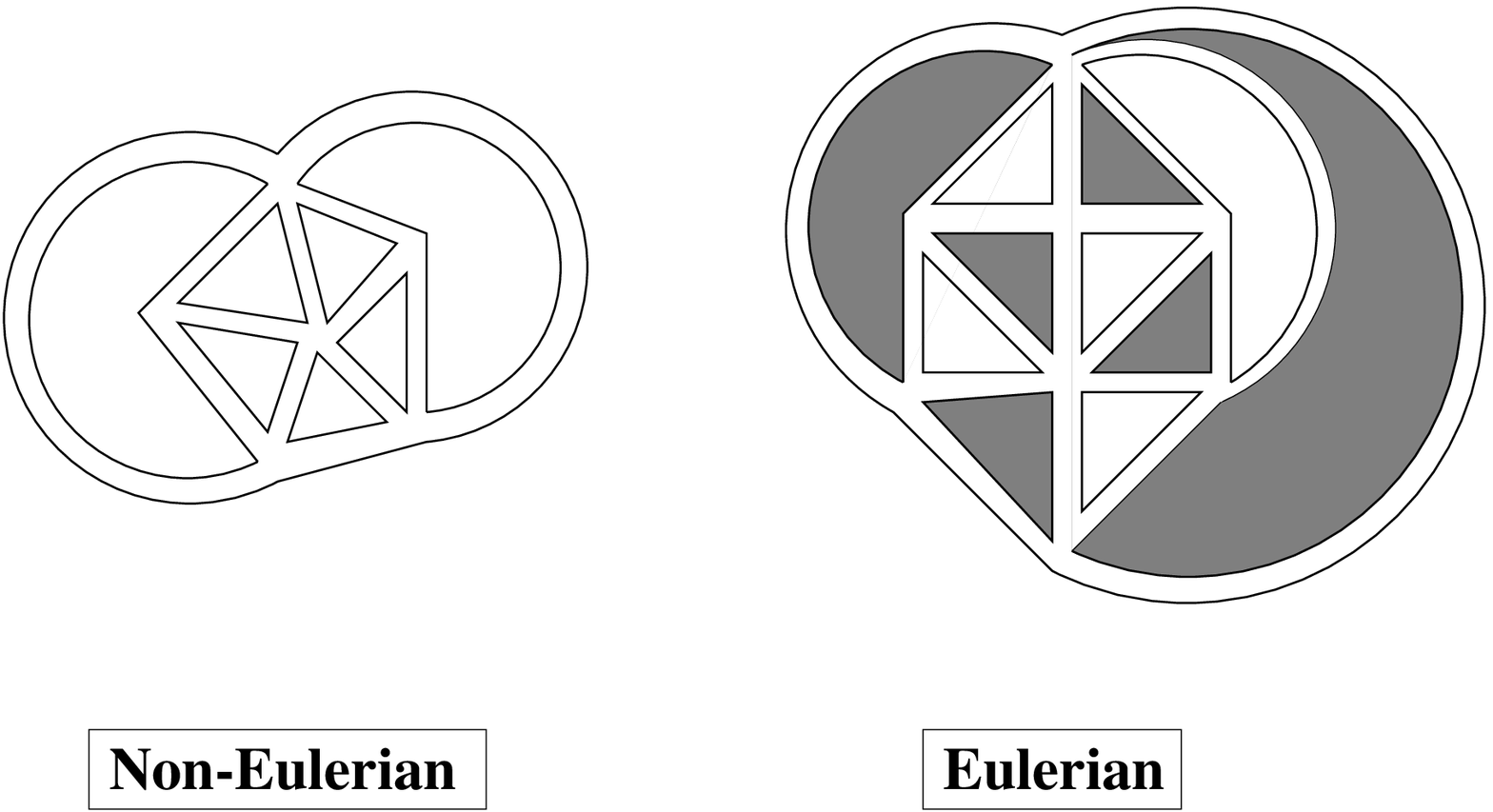}{10.cm}
\figlabel\eulernon

Let us go back for a while to the fully-packed loop model with $n=2$, interpreted
above as the model of phantom folding of the triangular lattice. 
Recall that in that case the height variable has been identified with the position in $\IR^2$
of the vertices in the folded configuration.
We wish now to define this
model on a {\it random triangulation}. For this to make sense, we need the triangulation to
be foldable. 
Indeed, if it is not, the height variable can no longer be defined as the (2D) position
in the folded configuration, but rather like in the dense case by viewing the loops as a contour
plot for the (1D) height.
In genus $0$, the foldability amounts to the fact that each vertex
is adjacent to an {\it even} number of triangles. In other words, the triangulation may
be {\it bicolored}. Such a triangulation is also called Eulerian, as this guarantees
the existence of an Eulerian path, visiting all triangles of the triangulation.
Sample Eulerian and non-Eulerian triangulations are depicted in Fig. \eulernon\ 
for illustration.
Summing over such triangulations will be called Eulerian gravity for obvious reasons.
Now we may define two possible ``gravitational" models of fully-packed loops on random trivalent
graphs.

$\bullet$ We may sum over {\bf arbitrary} trivalent graphs. As the dual triangulation will
in general {\it not} be Eulerian, it will not be foldable, and the extra degree of freedom
will be lost. The string susceptibility of the corresponding gravitational model
(with as usual a weight $g$ per trivalent vertex) will be computed using \kpz\ with the
dense central charge 
\eqn\orgra{ {\rm ordinary}\ {\rm gravity}: c=1-6 {e^2\over 1-e} , \quad n=2\cos\pi e}

$\bullet$ We may sum over {\bf bipartite} trivalent graphs, whose dual triangulation is
automatically foldable, thus preserving the height variable in $\IR^2$. The
string susceptibility must be computed using \kpz\ with the fully-packed central charge
\eqn\charfuly{ {\rm Eulerian} \ {\rm gravity}: c=2 -6 {e^2\over 1-e} , \quad n=2\cos\pi e}

So we expect here a pheomenon very similar to that occurring in the case of hard particles.
The universality class of the critical point depends crucially on the colorability
property of the underlying random lattices. This similarity is no coincidence, as the
very existence of {\it completely folded} groundstates relies crucially on the bipartiteness
of the lattice just like the crystalline groundstates of the hard-particle model did. 

To make the above argument even more solid, let us give a matrix model derivation of
the results \orgra\ and \charfuly\ in the case $n=1$.
The fully-packed model on random trivalent graphs is described by the partition function
\eqn\partorgra{ Z(g)= \int dA dB e^{-N{\rm Tr}({A^2\over 2}+{B^2\over 2}-g A B^2)} }
where as in Sect. 3.2 we use matrix elements of $A$ and $B$, two $N\times N$ Hermitian matrices,
to generate empty (resp. occupied) half-edges glued into empty (resp. occupied) edges
via the propagators $\langle A A\rangle$ and $\langle B B\rangle$.
Note that the measure of integration in \partorgra\ is normalized in such a way that $Z(g=0)=1$.
As now usual, Log$\, Z(g)$
generates a sum over connected trivalent fatgraphs decorated by fully-packed loops
of $B$ matrix elements. We have $n=1$, as there is no extra weight per loop.
Noting that the dependence of $Z(g)$ on $A$ is Gaussian, we may explicitly integrate
over $A$, with the result:
\eqn\resintegra{ Z(g)=\int dB e^{-N{\rm Tr}({B^2\over 2} -{2g^2} {B^4\over 4})} }
As we already mentioned in Sect. 3.2, this latter integral simply generates random 
{\it quadrangulations}, therefore is a model of pure gravity with no matter, i.e.
with $c=0$, corresponding to the dense formula \orgra, with $n=1$, $e=1/3$. 

To generate bipartite trivalent graphs (dual to Eulerian triangulations), we
must use {\it complex} matrices $A,B$, whose pictorial representation bears an extra 
orientation, to distinguish between $A$ and $A^\dagger$. The matrix integral
for the fully-packed loop model on bipartite trivalent graphs reads \DGK\
\eqn\intbipa{ Z_E(g)=\int dA dB e^{-N{\rm Tr}(A A^\dagger+B B^\dagger -g(AB^2+A^\dagger
(B^\dagger)^2)} }
where $A$ and $B$ are complex $N\times N$ matrices, and the measure is normalized in such a way
that $Z_E(g=0)=1$. We note that the dependence on $A$ is still Gaussian, hence upon
integrating it out, we get
\eqn\eulerfin{ Z_E(g)=\int dB e^{-N{\rm Tr}( B B^\dagger -g^2 B^2 (B^\dagger)^2 )} }
This turns out to be a particular case of the six-vertex model coupled to 2DQG solved
in \KZJ. The corresponding central charge is $c=1$, thus confirming
the formula \charfuly\ at $n=1$, $e=1/3$. 

The lesson to be drawn from this section is simple: the extra degree of freedom of the fully-packed
loop model is wiped out if we are not careful enough with the type of random lattices
we sum over. Only if these are bipartite will the height variable survive and lead eventually
to the central charge \charfuly. Otherwise, the mismatch between the random lattices and the model
takes the universality class back to that of the dense model. 
We expect this result to be quite general and we will apply it in the next section to the case
of the two-flavor fully-packed loop model on random tetravalent graphs.

\subsec{From fully-packed loops on the square lattice to meanders}

We are now ready to turn to the meander problem. It is by now clear that a meander configuration
is nothing but a configuration of the two-flavor loop model on a random tetravalent graph,
with in addition exactly {\it one} loop of each color, say the river (color $1$) and the road
(color $2$). This assumes that we have closed the river (infinite line)
into a loop as well. The operation of cutting the river loop and extending it into a line
is equivalent to a marking (rooting) of the corresponding diagram, and yields an overall factor
of $4 n$ for diagrams with $2n$ vertices. 
In the language of the two-flavor fully-packed loop model, extracting the configurations
with one loop of each color corresponds
to letting $n_1,n_2\to 0$, in order to select the leading contribution to the partition
function, proportional to $n_1n_2$ in this limit.  But why at all consider the FPL$^2$ model?
Indeed, only the ``crossing" vertex of Fig. \vertex\ (a) occurs in a meander configuration. 
The problem is that a two-loop flavor model on the square lattice with only the crossing vertex
allowed is trivial, as loops should all be straight lines (closed into loops say by periodic
boundary conditions). It simply means that the degrees of freedom of the lattice version
of the meander model are not taken into account by this naive definition. We will show that
the FPL$^2$ model provides the correct lattice version of the meander model.

\fig{A sample bipartite tangent meander with 8 vertices, among which 6 are crossing  vertices
(bridges) and 2 are tangency vertices. The vertices are bicolored, and all river
or road edges connect only vertices of distinct colors.}{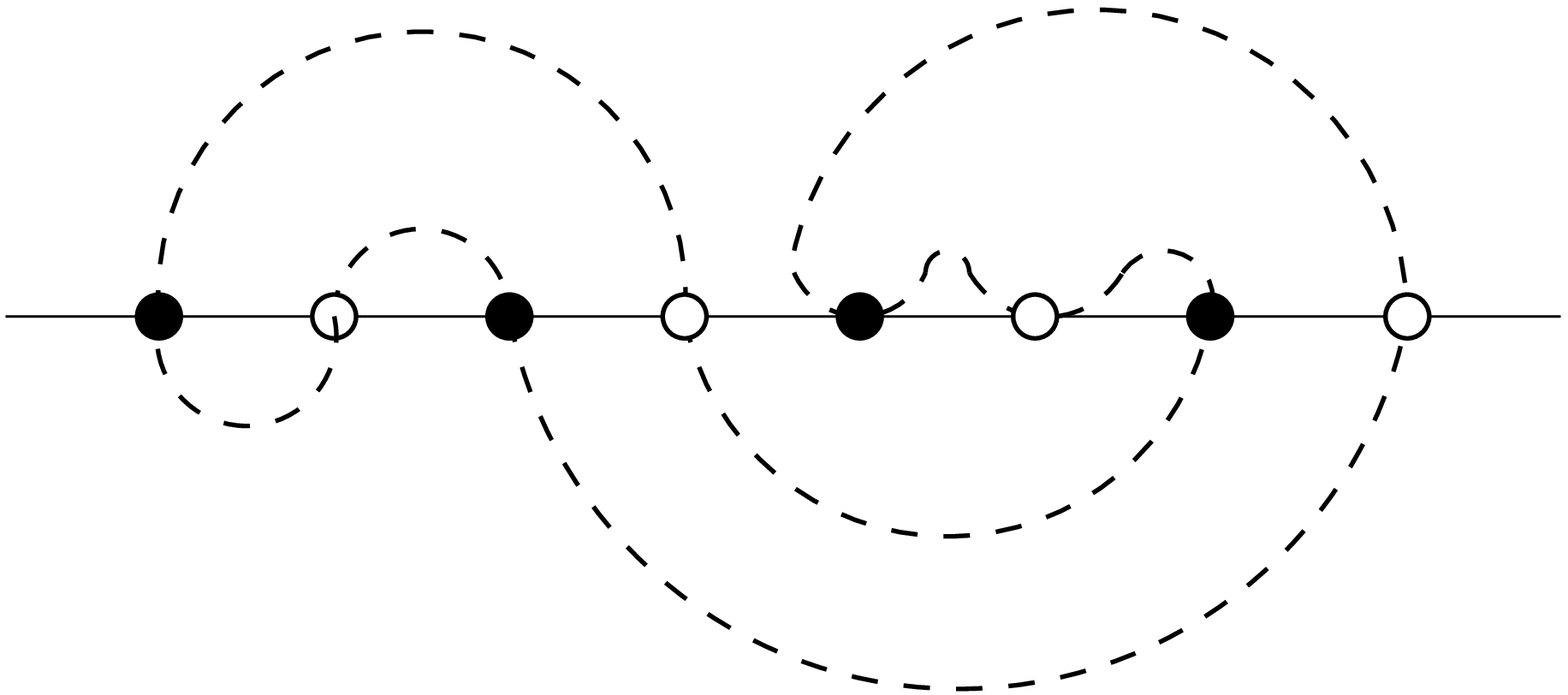}{6.cm}
\figlabel\bitangent

Let us now study the FPL$^2$ model coupled to 2DQG, namely defined on random tetravalent
graphs. 
As explained in Sect. 2, the genus zero partition function for this model reads
\eqn\pfgra{ Z(n_1,n_2;g,x,y)=\sum_{{\rm tetravalent}
\ {\rm planar}\atop {\rm graphs}\  \Gamma }
{1\over |{\rm Aut}(\Gamma)|} \sum_{{\rm FPL}\ {\rm configs.}\atop
{\rm on } \ \Gamma}  n_1^{L_1} n_2^{L_2} (gx)^{V_a(\Gamma)} (gy)^{V_b(\Gamma)} }
where the sum extends over all planar four-valent graphs $\Gamma$, and
$|$Aut$(\Gamma)|$ is the order of the
symmetry group of $\Gamma$, while we have also denoted by $V_a,V_b$
the total numbers of vertices of type a and b defined in Fig. \vertex\
in the particular loop configuration, namely we have weighted each crossing of a
black and a white
loop by $gx$ and each tangency by $gy$. From the lesson of Sect. 4.3, 
we know that if we wish to preserve all degrees
of freedom of the model, we must couple it to Eulerian gravity, namely consider
it on {\it bipartite} random tetravalent graphs in the sum \pfgra\ (we will distinguish this 
particular summation by an index $E$ for Eulerian in the notation for the partition function). 
Indeed, recall that the definition of
the (3D) height variable in Sect. 4.2 has relied crucially on the preliminary bicoloration
of vertices of the lattice, allowing for defining the edge variables $A,B,C,D$.
We may therefore use \pfgra\ to generate ``bipartite tangent meanders", 
namely configurations of a closed road
crossing or tangent to a river (infinite line) through a given number of bicolored points $2n$,
such that only points of distinct color are connected by river or road edges.
Let $\mu_{2n}$ denote the number of bipartite tangent meanders with $2n$ vertices 
(see Fig. \bitangent\ 
for an illustration). These are well
enumerated by counting configurations of the FPL$^2$ model at $n_1=n_2=0$, namely $e_1=e_2=1/2$,
and by taking $x=y=1$ in \pfgra. More precisely, we have
\eqn\pftgmean{ \lim_{n_1,n_2\to 0}{1\over n_1n_2}
(Z_E(n_1,n_2;g,1,1)-1)=\sum_{n\geq 1} {\mu_{2n} \over 4n} g^{2n} } 
where we have expressed that the opening of the river into an infinite line
amounts to the marking of a river edge, and results in a factor $4n$ between the 
fixed $n$ free energy $f_{2n}$ and the tangent meander number $\mu_{2n}$.
Plugging the values $e_1=e_2=1/2$ into the formula \cenchartwo\ and using \kpz, we find
\eqn\restang{ c=-3 \qquad \gamma_{str}= -{1+\sqrt{7}\over 3} }
The free energy at fixed area $A$ for this model is nothing but 
\eqn\fremeant{ F_A = {\mu_A \over 2A} \sim {g_c^{-A} \over A^{3-\gamma_{str}} }}
by directly applying \largA.
Note that the value of $g_c$ is not predicted by this argument, only the configuration
exponent. 
We finally deduce the large $n$ asymptotics of the bipartite tangent meander numbers
\eqn\asymu{ \mu_{2n} \sim {g_c^{-2n} \over n^{\beta} } \qquad \beta={7+\sqrt{7} \over 3}  }

\fig{Typical height configurations around (a) a crossing vertex and 
(b) a tangency vertex.
When going diagonally say from the NW to the SE face (shaded), the height increases
by an amount restricted to $\pm ({\bf A}-{\bf C}),\pm ({\bf B}+{\bf C})$
in the case (a)
and only to $\pm ({\bf A}+{\bf B})$ in the case (b). Forbiding (b)
will therefore amount to a reduction of the height variable range from 3 to 2 
dimensions.}{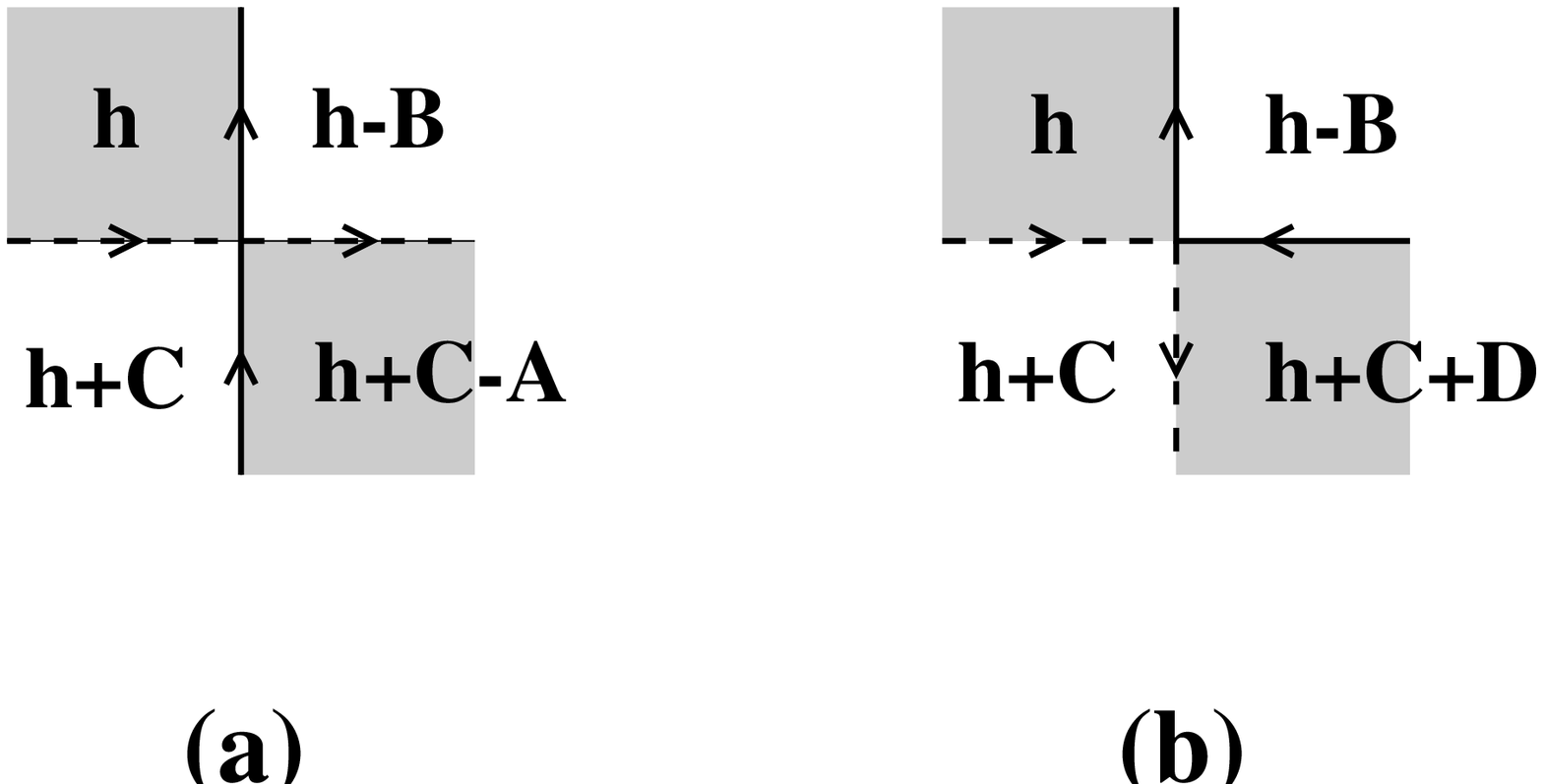}{9.cm}
\figlabel\forbid

This is not however the end of the story. Indeed, the numbers $\mu_{2n}$ might have
nothing to do with the meander numbers $M_{2n}$. We must examine more closely the role
of the tangency points.
Note that the meander numbers are correctly generated by taking $n_1,n_2,y\to 0$
in \pfgra:
\eqn\meagen{ \lim_{n_1,n_2\to 0}{1\over n_1n_2}
(Z_E(n_1,n_2;g,1,0)-1)=\sum_{n\geq 1} {M_{2n} \over 4n} g^{2n} }
which just amounts to forbiding the tangency vertex b of Fig. \vertex.
Forbiding the tangency vertex b will further restrict the range of the height
variable of the model. In Fig. \forbid, we have displayed typical height configurations
around a type a and a type b vertex for comparison. It is easy to see that the increase
in height when going diagonally from a face to another 
through the vertex 
(shaded faces in Fig. \forbid), 
there is more freedom at the crossing vertex a, where we may have
$h \to h\pm ({\bf A}-{\bf C})$ or $h \pm ({\bf B}+{\bf C})$, while at the tangent
vertex $b$ we may only go in one spatial direction $h \to h\pm ({\bf A}+{\bf B})$.
Forbiding the b vertex therefore has the net effect of dimensionally reducing the
range of the height variable to only {\it two} dimensions rather than three in the
full model. Forbiding tangencies therefore takes the central charge back to the dense value,
which for $e_1=e_2=1/2$ reads
\eqn\denseval{ c=-4  \qquad \Rightarrow \qquad \gamma_{str}=-{5+\sqrt{145}\over 12} }
Finally, using \meagen\ and applying \largA, we get the following meander 
number asymptotics for large $n$  
\eqn\measympto{ M_{2n} \sim {g_c^{-2n} \over n^{\alpha}} \qquad \alpha={29+\sqrt{145} 
\over 12}  }

To make the above argument more solid, let us present an alternative route leading to
the result \denseval. We may have considered in \pfgra\ a summation over
arbitrary (non-necessarily bipartite) tetravalent graphs. In this case, the
bicoloration of vertices disappears, and we have therefore no way of 
distinguishing the vector labels $\bf A$ and $\bf -B$ on one hand and
$\bf C$ and $\bf -D$ on the other hand. This amounts to imposing
an extra condition 
\eqn\densecond{ {\bf A}+{\bf B}={\bf C}+{\bf D}=0 }
This is precisely the condition one would obtain by considering the dense
two-flavor loop model on the square lattice. Indeed, we still have the dictionary
of Fig. \dictio\ for edge labels, but as more vertices are allowed, we
get some extra conditions for the height variable to be well-defined. If we consider a
vertex crossed by only a black or white loop we immediately see that \densecond\
must be satisfied. This explains the reduction $3\to 2$ in the expression for
the dense central charge \cenchartwo\ compared to the fully-packed one, as the (2D) height
is now a linear combination of ${\bf A},{\bf C}$ which we may choose 
to form an orthonormal basis of $\IR^2$. 
Hence the simple fact that we sum over arbitrary tetravalent graphs takes
the universality class of the FPL$^2$ model back to that of the dense one.
For $e_1=e_2=1/2$, we find precisely the result \denseval.

\fig{A type b vertex of the FPL$^2$ model coupled to
ordinary 2DQG, together
with its dual height configuration. We note that the NE and SW heights
are identical. We may therefore undo the vertex as shown, which explains
its irrelevance.}{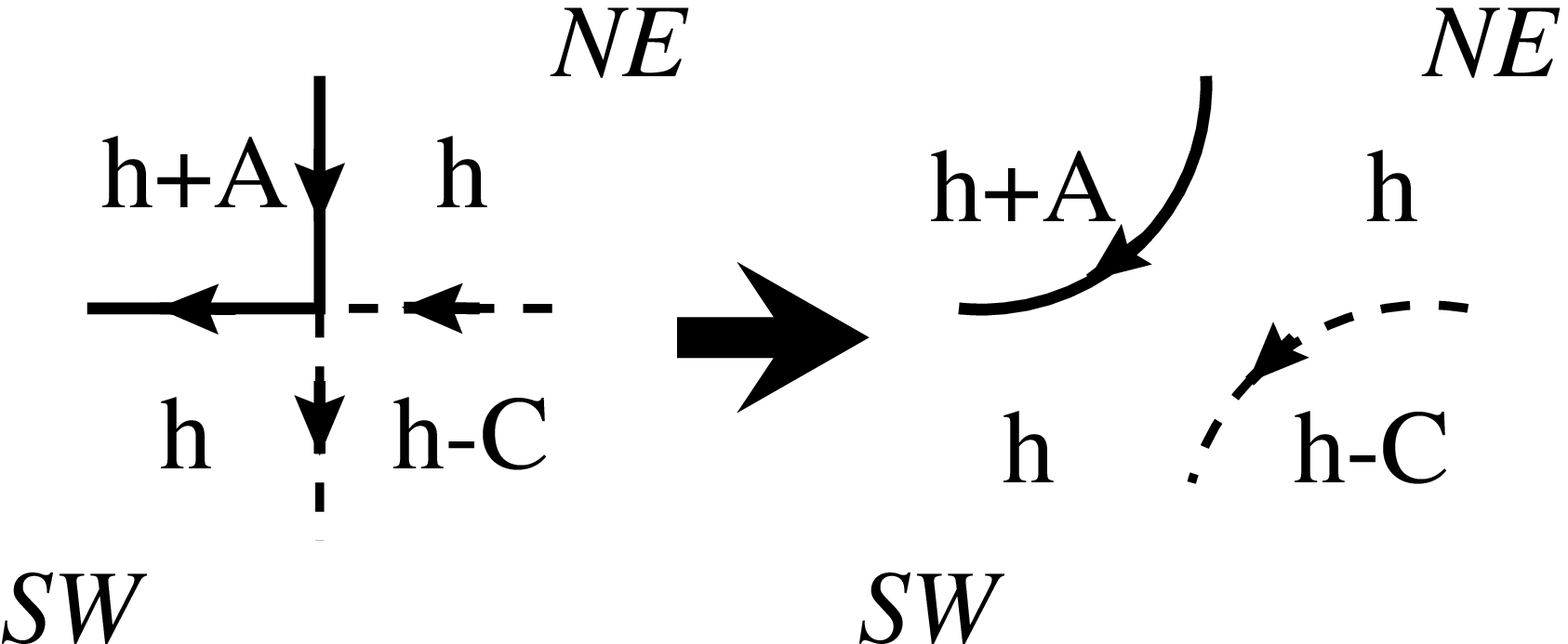}{7.cm}
\figlabel\irrel

We must now see where meanders fit into this picture. Recall that we still have
at this point crossing and tangent vertices, but we have relaxed the constraint
of bicolorability. We must again examine more closely the role
of the tangency points. Forbiding them will yield the meander configurations as
\eqn\viamean{ \lim_{n_1,n_2\to 0}{1\over n_1n_2}
(Z(n_1,n_2;g,1,0)-1) = \sum_{n\geq 1} {M_{2n} \over 4n} g^{2n} }
Examining the height variable of the dense model
around a tangency vertex, we see that the height variable
is the same in the faces lying outside of the two loops (see Fig. \irrel).
This means simply that the tangency vertex is {\it irrelevant} in the ordinary
gravity situation, namely that its presence or absence, although it might
affect the precise value of $g_c$, doesn't affect the universality class,
hence the configuration exponent is independent of $y$ at $x=1$ in \pfgra.
We therefore end up with the meander universality class, determined by
\densecond.   

Some important numerical checks of the results of this section have been carried out
by use of Jensen's transfer matrix method \JEN, which consists in generating
the meander or tangent meander or bipartite tangent meander configurations  
by applying a local ``transfer matrix" along the river, which implements
the addition of a new vertex to a previous state. The method is powerful enough
to give very accurate values for the extrapolated configuration exponents,
which all confirm the various results obtained above \DGJ. Note that these are 
somewhat contested in \POLEM, however the discrepancy with the predicted
exponents is extremely small and possibly due to subtleties with the large
$n$ corrections.

\subsec{The meander universality class: asymptotics of meandric numbers}

We have now identified the CFT underlying meanders, as the dense two-flavor
loop model
with $n_1=n_2=0$ coupled to ordinary 2DQG. The complete knowledge of the 
conformal operator content of this CFT via the Coulomb gas picture  
gives access to a host of meandric numbers which we describe now. 

The important operators for our present purpose are those identified as generating
oriented river vertices, namely the operators $\phi_{k}$ (resp. $\phi_{-k}$), 
$k=1,2,...$ which
correspond to the insertion of a $k$-valent source (resp. sink)
vertex from (resp. to) which $k$ 
oriented river edges originate (resp. terminate). Using these,
we may generate configurations in which the river may
itself form a complicated though connected oriented planar graph with possibly 
loops and endpoints, either sources or sinks according to the operators.
In the Coulomb gas picture, these operators create ``magnetic" defect lines
along which the height variable has discontinuities.
The operator $\phi_k$ has conformal dimension \JACO\
\eqn\dimpsi{ h_k = {k^2-4 \over 32} }
$k=\pm1,\pm2,...$
When coupled to 2DQG, these operators get dressed (into $\Psi_k$)
and acquire the dimension
\kpzdim:
\eqn\acqdim{ \Delta_k={{1\over 2}\sqrt{8+3 k^2}- \sqrt{5}\over \sqrt{29}-\sqrt{5}} }

As a preliminary remark, we note that $h_{\pm 2}=\Delta_{\pm 2}=0$.
$\Psi_{\pm 2}$ indeed corresponds to the marking of an edge of the river in meanders,
and moreover such operators must go by source/sink pairs for the orientations of the pieces
of river connecting them to be compatible.
Applying \corrA\ to the two-point correlator $\langle \Psi_2 \Psi_{-2}\rangle_A$ 
at fixed large area $A$, we find
\eqn\markpoi{ \langle \Psi_2 \Psi_{-2}\rangle_A \sim {g_c^{-A} \over A^{1-\gamma_{str}} }}
while the meander counterpart (with a closed river) behaves as
$M_A/(2A)\sim g_c^{-A}/(A^{3-\gamma_{str}})$. We see that the net effect of the insertion
of the operators $\Psi_{\pm 2}$ is an overall factor 
proportional to $A^2$, which confirms their interpretation
as marking operators. These are very important to keep in mind, as they might be
required to ensure source/sink balance in various river geometries.

We may now turn to the case of semi-meanders, namely meanders in which the river is a 
semi-infinite line around the origin of which the road may freely wind. 
Considering the point at infinity on the river as just another point, the semi-meanders
may equivalently be viewed as meanders whose river is made of a segment. Sending
one of the ends of the segment to infinity just resolves the winding ambiguities around
both ends. Using the above river insertion operators, we immediately identify
the generating function for semi-meanders as
\eqn\serexP{ \langle \Psi_1 \Psi_{-1}\rangle = \sum_{n\geq 1} {\bar M}_n g^n}
Using again \corrA\ and the explicit values of $\Delta_{\pm 1}$ via \acqdim, 
we arrive at the large $n$ asymptotics
\eqn\larnsem{ {\bar M}_n\sim  {{g_c}^{-n}\over n^{\bar \alpha}}
\qquad {\bar \alpha}=1+2 \Delta_1-\gamma_{str}= 1+{\sqrt{11}\over 24}(\sqrt{5}+\sqrt{29}) }
Note that we expect the value of $g_c$ to be the {\it same} for
meanders and semi-meanders, as both objects occur as thermodynamic quantities in the {\it
same} effective field theory.

\fig{Three types of meandric configurations in which the river has the geometry of (a)
a $k$-valent star (b) an ``eight" (c) a ``cherry". The vertices corresponding to 
river sinks or sources are represented by filled circles ($\bullet$). The edges
of river inbetween them are oriented accordingly. The road (dashed line)
may freely wind around univalent vertices.}{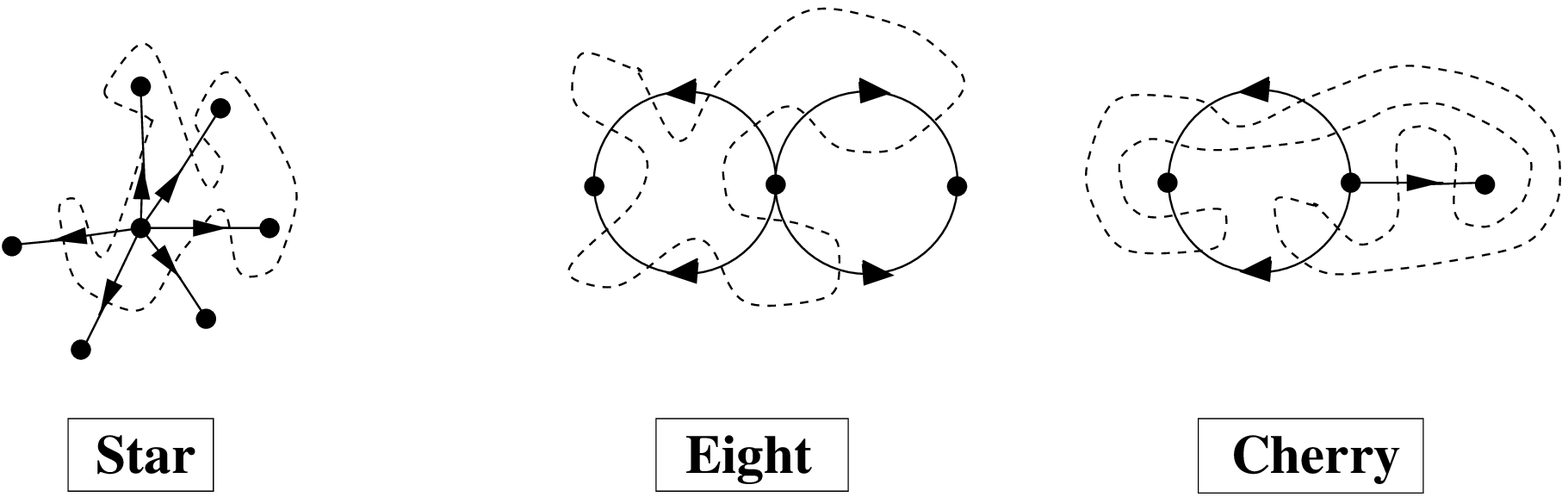}{12.cm}
\figlabel\meandric

We may now generate many more meandric numbers by considering more general correlators.
To name a few (all depicted in Fig. \meandric), 
we may generate rivers with the geometry of a star with one
$k$-valent source vertex and $k$ univalent sink vertices generated
by $\langle \Psi_k (\Psi_{-1})^k\rangle$, rivers with the geometry of an ``eight"
with one tetravalent source vertex and two loops, each containing a bivalent sink vertex
generated by $\langle \Psi_4 (\Psi_{-2})^2\rangle$, or rivers with the geometry of
a ``cherry" with one trivalent source vertex, one univalent sink, and one loop,
marked by a bivalent sink vertex generated by 
$\langle \Psi_3 \Psi_{-1}\Psi_{-2}\rangle$, etc ... For each of these situations,
we get the corresponding configuration exponent 
$\alpha=3-\gamma+\sum_i (\Delta_i-1)$ by applying \corrA\ with the dimensions
\acqdim.
We get respectively
\eqn\respget{\eqalign{
\alpha_{\rm k-star}&={1\over 48}(\sqrt{5}+\sqrt{29})(\sqrt{3k^2+8} +k(\sqrt{11}-2 \sqrt{29})
+4 \sqrt{29}-2\sqrt{5}) \cr
\alpha_{\rm eight}&={1\over 24}(\sqrt{5}+\sqrt{29})(\sqrt{14}+\sqrt{5})\cr
\alpha_{\rm cherry}&=
{1\over 48}(\sqrt{5}+\sqrt{29})(\sqrt{11}+\sqrt{35}) \cr }}
These values were also checked numerically in \DGJ.

\newsec{Conclusion}

In these notes, we have tried to clarify the role of random lattices when coupling
geometrically constrained systems such as hard particles or fully-packed loops to 
2DQG. We have in both cases shown that the application of the
famous KPZ formulas \kpz\-\kpzdim\ can be subtle, and involves the correct
understanding of the models' degrees of freedom both on the fixed and random
lattices. The main lesson is that if we want to preserve
the essence (universality class) of a given model when coupled to 2DQG, then
we must pay the price of restricting the fluctuations of space (random lattices)
to those and only those preserving these degrees of freedom.   

Provided we incorporate this lesson, we are free to use back and forth the KPZ
formulas to transpose fixed lattice results into random lattice ones and vice
versa. In the case of hard objects, we have used the 2DQG picture to infer some
result on fixed lattices. In the case of fully-packed loops, we have used the 
opposite strategy to solve the meander asymptotics problem by using fixed
lattice results on the dense or fully-packed two-flavor loop model. 
Along the way we have presented a number of other relations between fixed
and random lattice models.

One could still wonder, now that we know the exact answers, 
whether one could devise some more rigorous mathematical proof of the various
results inferred. In particular, the hard-triangle and hard-square
models remain to be solved. As to meanders, the existence of a transfer matrix
formulation may give some hope to be able to attain a combinatorial proof
of our assertions. One could also wonder whether the recent combinatorial
techniques for tackling planar graphs by use of trees might apply to meandric
or loop problems (see \SCH\ \PLA\ \HARDOB\ \CONCUR\ for instance). 

\fig{A configuration of the one-flavor fully-packed loop model
on a bipartite trivalent planar graph. We have opened the loop into a
straight line and represented the unoccupied edges as dashed non-intersecting arches. 
Both loop and non-loop edges connect only
vertices of distinct colors.}{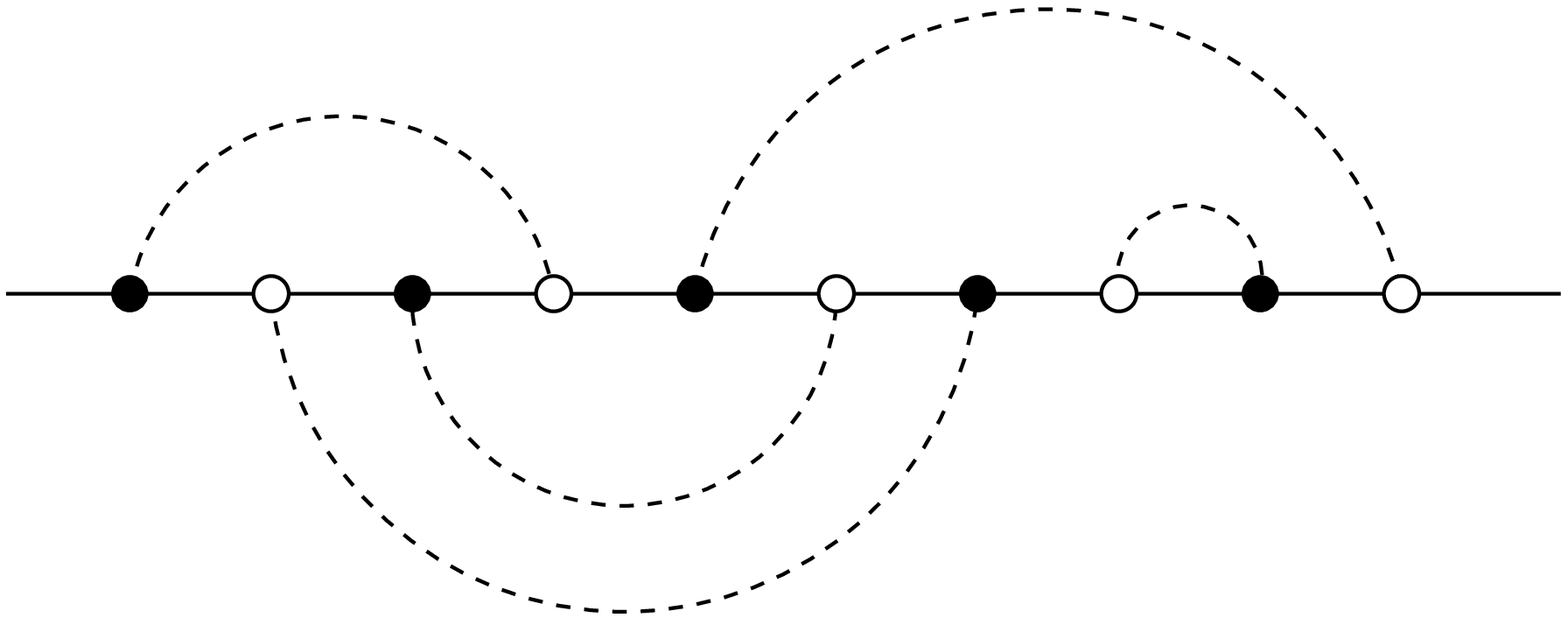}{9.cm}
\figlabel\hamil

As a final question, let us mention the following simpler enumeration problem
for which we also know the exact configuration exponent. It corresponds
to the one-flavor fully-packed loop model coupled to Eulerian 2DQG, when $n\to 0$
\GKN.
The corresponding configurations are simply trivalent planar graphs with bicolored
vertices and exactly one loop visiting all vertices. Upon opening the loop
into an infinite line, we must enumerate planar configurations of 
non-intersecting arches connecting bicolored
vertices on the line by pairs, in such a way that the colors of the connected vertices 
are always distinct (both along the river and along arches). A sample such configuration
is depicted in Fig. \hamil. 
The resulting number of configurations for a fixed number $2n$
of vertices, $\nu_{2n}$, 
counts also by duality the number of Eulerian triangulations with a Hamiltonian cycle.
This number has the following large $n$ asymptotics   
\eqn\larnu{ \nu_{2n} \sim {g_c^{-2n}\over n^{\delta} }\qquad \delta={13+\sqrt{13}\over 6} }
for some constant $g_c$.
This is obtained by applying \largA\ with the string susceptibility \kpz\ of the
fully-packed loop model with $n=0$ coupled to Eulerian gravity, namely
with the fully-packed central charge \cchar\ with $e=1/2$, hence $c=-1$, and 
$\gamma_{str}=-{1+\sqrt{13}\over 6}$.
Challenge: prove eqn. \larnu!

\listrefs
\end